\RequirePackage{lineno}
\documentclass[prd,amsmath,amssymb,showpacs,superscriptaddress,nofootinbib]{revtex4}
\usepackage{amssymb}
\usepackage{multirow}
\usepackage{graphicx,epsfig}
\usepackage{dcolumn}
\usepackage{bm}
\usepackage[dvipdfm,CJKbookmarks=true,unicode,colorlinks,linkcolor=blue,anchorcolor=blue,citecolor=blue,pdfborder={0 0 0}]{hyperref}
\usepackage{amsfonts}
\usepackage{keyval,graphicx}
\usepackage{textcomp,wasysym}
\usepackage{def-com}

\begin{document}
\parskip=3pt plus 1pt minus 1pt

\title{\boldmath Study of the near-threshold $\omega\phi$ mass enhancement
in doubly OZI suppressed $\jpsitogwf$ decays}
\author{
\small
M.~Ablikim$^{1}$, M.~N.~Achasov$^{6}$, O.~Albayrak$^{3}$, D.~J.~Ambrose$^{39}$,
F.~F.~An$^{1}$, Q.~An$^{40}$, J.~Z.~Bai$^{1}$, Y.~Ban$^{26}$, J.~Becker$^{2}$,
J.~V.~Bennett$^{16}$
% ...
\\
\vspace{0.2cm}
(BESIII Collaboration)\\
\vspace{0.2cm} {\it
$^{1}$ Institute of High Energy Physics, Beijing 100049, People's Republic of China\\
$^{2}$ Bochum Ruhr-University, D-44780 Bochum, Germany\\
$^{3}$ Carnegie Mellon University, Pittsburgh, Pennsylvania 15213, USA\\
% ...
$^{6}$ G.I. Budker Institute of Nuclear Physics SB RAS (BINP), Novosibirsk 630090, Russia\\
% ...
$^{16}$ Indiana University, Bloomington, Indiana 47405, USA\\
% ...
$^{26}$ Peking University, Beijing 100871, People's Republic of China\\
% ...
$^{39}$ University of Rochester, Rochester, New York 14627, USA\\
$^{40}$ University of Science and Technology of China, Hefei 230026, People's Republic of China\\
% ...
}
\vspace{0.4cm}
}


\begin{abstract}
A 2.25$\times$10$^8$ $\jpsi$ event sample accumulated with the BESIII detector is used to study
the doubly OZI suppressed decay modes $\jpsi\to\gamma\of$, $\omega\to\ppp$,  $\phi\to\kk$.
A strong deviation ($>$ 30$\sigma$) from three-body $\jpsi\to\gamma\omega\phi$ phase space
is observed near the $\omega\phi$ mass threshold that is consistent with a previous
observation reported by the BESII experiment.
A partial wave analysis with a tensor covariant amplitude that assumes
that the enhancement is due to the presence of a resonance, the $X(1810)$, is performed,
and confirms that the spin-parity of the $X(1810)$ is $0^{++}$.
The mass and width of the $X(1810)$ are determined to be
$M=1795\pm7$(stat)$^{+13}_{-5}$(syst)$\pm$19(mod) MeV/$c^2$ and
$\Gamma=95\pm10$(stat)$^{+21}_{-34}$(syst)$\pm$75(mod) MeV/$c^2$, respectively, and the product branching fraction is measured to be
${\cal B}(\jpsi\to\gamma X(1810))\times{\cal B}(X(1810)\to\of)=(2.00\pm0.08$(stat)$^{+0.45}_{-1.00}$(syst)$\pm$1.30(mod))$\times10^{-4}$.
These results are consistent within errors with those of the BESII experiment.
\end{abstract}


\maketitle

\section{Introduction}
An anomalous near-threshold enhancement in the $\of$ invariant-mass spectrum
in the process $\jpsitogwf$ was reported by the BESII experiment~\cite{Ablikim:2006}.
A partial wave analysis (PWA) that used a helicity covariant amplitude that assumed that the enhancement was produced by a resonance, denoted as the $X(1810)$,
was performed on the BESII event sample.
The analysis indicated that the $X(1810)$ quantum number assignment
favored $J^{PC}=0^{++}$ over $J^{PC}=0^{-+}$ or $2^{++}$ with
a significance of more than 10$\sigma$. The mass and width were determined to be
$M = 1812^{+19}_{-26}$(stat)$\pm18$(syst) MeV/$c^2$ and
$\Gamma = 105\pm20$(stat)$\pm28$(syst) MeV/$c^2$, respectively, and the product branching
fraction ${\cal B}$($\jpsi\to\gamma$ $X(1810)$) $\cdot$ ${\cal B}$($X(1810)$$\to\wf$)
=$[2.61\pm0.27$(stat)$\pm0.65$(syst)]$\times10^{-4}$ was measured.
The $\jpsitogwf$ decay mode is a doubly OZI suppressed process with a
production rate that is expected to be suppressed relative to $\jpsi\to\gamma\omega\omega$ or $\jpsi\to\gamma\phi\phi$
by at least one order of magnitude~\cite{oneorder}.
Possible interpretations of the $\omega\phi$ threshold enhancement include a new type of resonance,
such as a tetraquark state (with structure $q^2\overline{q}^2$)~\cite{Bing-An:2006}, a hybrid~\cite{Kung-Ta:2006},
or a glueball state~\cite{Bicudo:2007} etc., a dynamical effect arising from intermediate meson rescattering~\cite{Qiang:2006},
or a threshold cusp of an attracting resonance~\cite{D.V.:2006}.
As of now none of these interpretations has either been established or ruled out by experiment.

A search for the $X(1810)$ was performed by the Belle
collaboration in the decay of $B^{\pm}\to K^{\pm}\omega\phi$~\cite{belle},
but no obvious $X(1810)$ signal was observed.
A high statistics data sample collected with the BESIII detector provides
a good opportunity to confirm the existence of the $\omega\phi$ threshold enhancement,
study its properties and search for other possible related states that decay to $\of$.

In this paper we present a PWA that uses a tensor covariant amplitude for the
$\jpsitogwf$ process, where the $\phi$ is reconstructed from $\kk$ and the $\omega$
from $\ppp$. The analysis is based on a sample of $(225.3\pm2.8) \times
10^{6} \jpsi$ events~\cite{jpsinumber} accumulated with the new Beijing
Spectrometer (BESIII)~\cite{bes3} located at the Beijing Electron-Positron
Collider (BEPCII)~\cite{bepc2}.

\section{Detector setup and Monte Carlo simulation}

BEPCII is a double-ring $e^{+}e^{-}$ collider designed to provide a
peak luminosity of $10^{33}$ cm$^{-2}s^{-1}$ with beam currents of
$0.93$~A. The BESIII detector has a geometrical acceptance of
$93\%$ of $4\pi$ and has four main components: (1) A small-cell,
helium-based ($40\%$ He, $60\%$ C$_{3}$H$_{8}$) main drift chamber
(MDC) with $43$ layers providing an average single-hit resolution
of $135$~$\mu$m, charged-particle momentum resolution in a $1$~T
magnetic field of $0.5\%$ at 1~GeV$/c$, and a $dE/dx$ resolution
better than $6\%$. (2) An electromagnetic
calorimeter (EMC) consisting of $6240$ CsI(Tl) crystals in a
cylindrical structure (barrel) and two endcaps. The energy
resolution for $1.0$~GeV$/c$ $\gamma$-rays is $2.5\%$ ($5\%$) in the barrel
(endcaps), and the position resolution is $6$~mm ($9$~mm) in the
barrel (endcaps). (3) A time-of-flight system (TOF) constructed of
$5$~cm thick plastic scintillators, with $176$ detectors of $2.4$~m
length in two layers in the barrel and $96$ fan-shaped detectors
in the endcaps. The barrel (endcap) time resolution of $80$~ps ($110$~ps)
provides $2\sigma$ $K/\pi$ separation for momenta up to $\sim 1.0$~GeV$/c$.
(4) The muon system (MUC) consists of $1000$~m$^{2}$ of Resistive
Plate Chambers (RPCs) in nine barrel and eight endcap layers and
provides $2$~cm position resolution.

In this analysis, a GEANT4-based~\cite{geant4} Monte Carlo (MC) simulation
software package, BOOST~\cite{boost}, is used. It provides an event
generator, contains the detector geometry description,
and simulates the detector response and signal digitization.
The production of the $\jpsi$ resonance is simulated by the Monte Carlo
event generator KKMC~\cite{kkmc1,kkmc2}, while the decays are generated
by BesEvtGen~\cite{besevtgen1,besevtgen2} for known decay modes with branching
ratios set at the PDG~\cite{PDG} world average values,
and by the Lund-Charm model~\cite{lundcharm} for the remaining unknown decays.
The analysis is performed in the framework of the BESIII
Offline Software System (BOSS), which takes care of
the detector calibration, event reconstruction, and data
storage.


\section{Event selection}

Signal $\jpsitogwf$ events with $\phitokk$ and $\omegatoppp$ final states
have the topology $3\gkkpp$. The event candidates are required to have
four well reconstructed charged tracks with net charge zero, and at
least three photons.

Charged-particle tracks in the polar angle range $|\costht|<0.93$ are
reconstructed from the MDC hits. Only the tracks with points of closest
approach to the beamline that are within $\pm$10 cm of the interaction point in
the beam direction, and within 1 cm in the plane perpendicular to the
beam are selected. TOF and $dE/dx$ information are combined to form
particle identification confidence levels for $\pi$, $K$ and $p$
hypotheses. Kaons are identified by requiring
the particle identification probability ($Prob$) to be
$Prob(K)>Prob(\pi)$ and $Prob(K)>Prob(p)$.
Two identified kaons with opposite charges are
required.

Photon candidates are reconstructed by clustering signals in EMC crystals.
The energy deposited in the nearby TOF counters is included to improve the
photon reconstruction efficiency and its energy resolution. The photon candidates
are required to be in the barrel region ($|\costht|<0.8$) of the EMC with at
least 25 MeV total energy deposition, or in the endcap regions ($0.86<|\costht|<0.92$)
with at least 50 MeV total energy deposition, where $\theta$ is the polar angle
of the shower. The photon candidates are furthermore required to be isolated
from all charged tracks by an angle $>$ 10$^\circ$ to suppress showers generated by charged
particles. The showers in the region between the barrel and the endcaps of the
EMC are poorly measured and excluded. Timing information from the EMC is used
to suppress electronic noise and energy deposits that are unrelated to the event.
Events with at least three good photon candidates are selected.

A four-constraint (4C) energy-momentum conserving kinematic fit is performed
to the 3$\gkkpp$ hypothesis. For events with more than three photon candidates,
the candidate combination with the minimum $\chi^2_{4C}$ is selected,
and it is required that $\chi^2_{4C}<40$ (the requirement is determined
by optimizing $S$/$\sqrt{S+B}$, where
$S$ is the number of MC signal events generated with phase space, and $(S+B)$ is the number of
signal plus background candidate events in the data).
In order to remove background stemming from $\jpsi\to$2$\gkkpp$ and $\jpsi\to$4$\gkkpp$,
we performed 4C kinematic fits for the hypotheses of 2$\gkkpp$ and 4$\gkkpp$ (for the
events that have at least four good photon candidates).
We require $\chi^{2}_{4C}(3\gkkpp)<\chi^{2}_{4C}(2\gkkpp)$ and
$\chi^{2}_{4C}(3\gkkpp)<\chi^{2}_{4C}(4\gkkpp)$, respectively.
The $\pi^0$ candidates are reconstructed from the two of the three selected
photons with invariant mass closest to the $\pi^0$ mass,
and $|M_{\GG}-M_{\piz}|<20$ MeV/$c^{2}$ is required.

\begin{figure}[htbp]
\vskip -0.1 cm \centering
{\includegraphics[width=5.8cm,height=6cm]{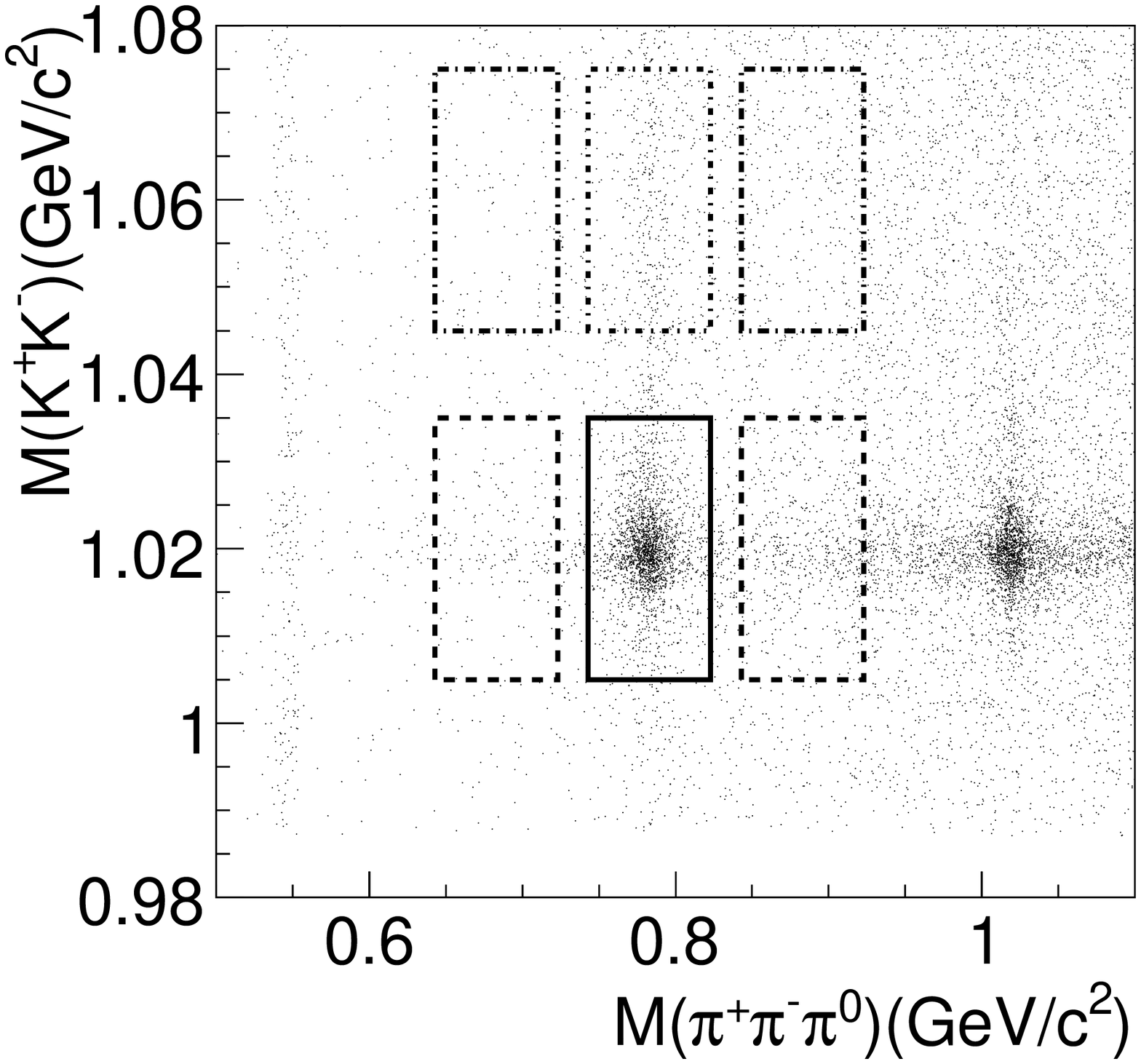}
  \put(-30,150){(a)}
  \put(-97,130){C}
  \put(-75,130){B}
  \put(-55,130){C}
  \put(-97,80){A}
  \put(-75,80){S}
  \put(-55,80){A}
\includegraphics[width=5.8cm,height=6cm]{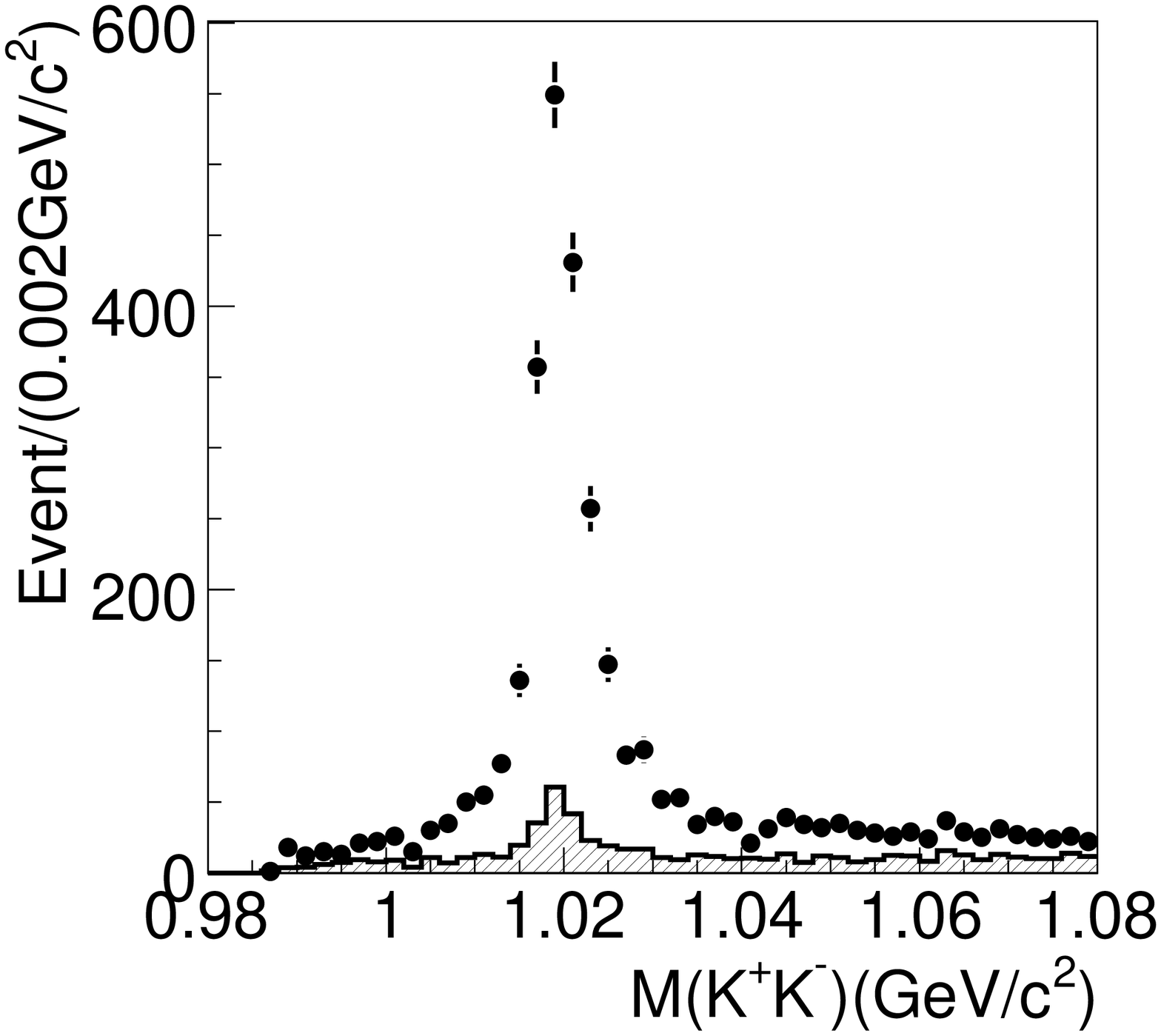}
\put(-25,150){(b)}
\includegraphics[width=5.8cm,height=6cm]{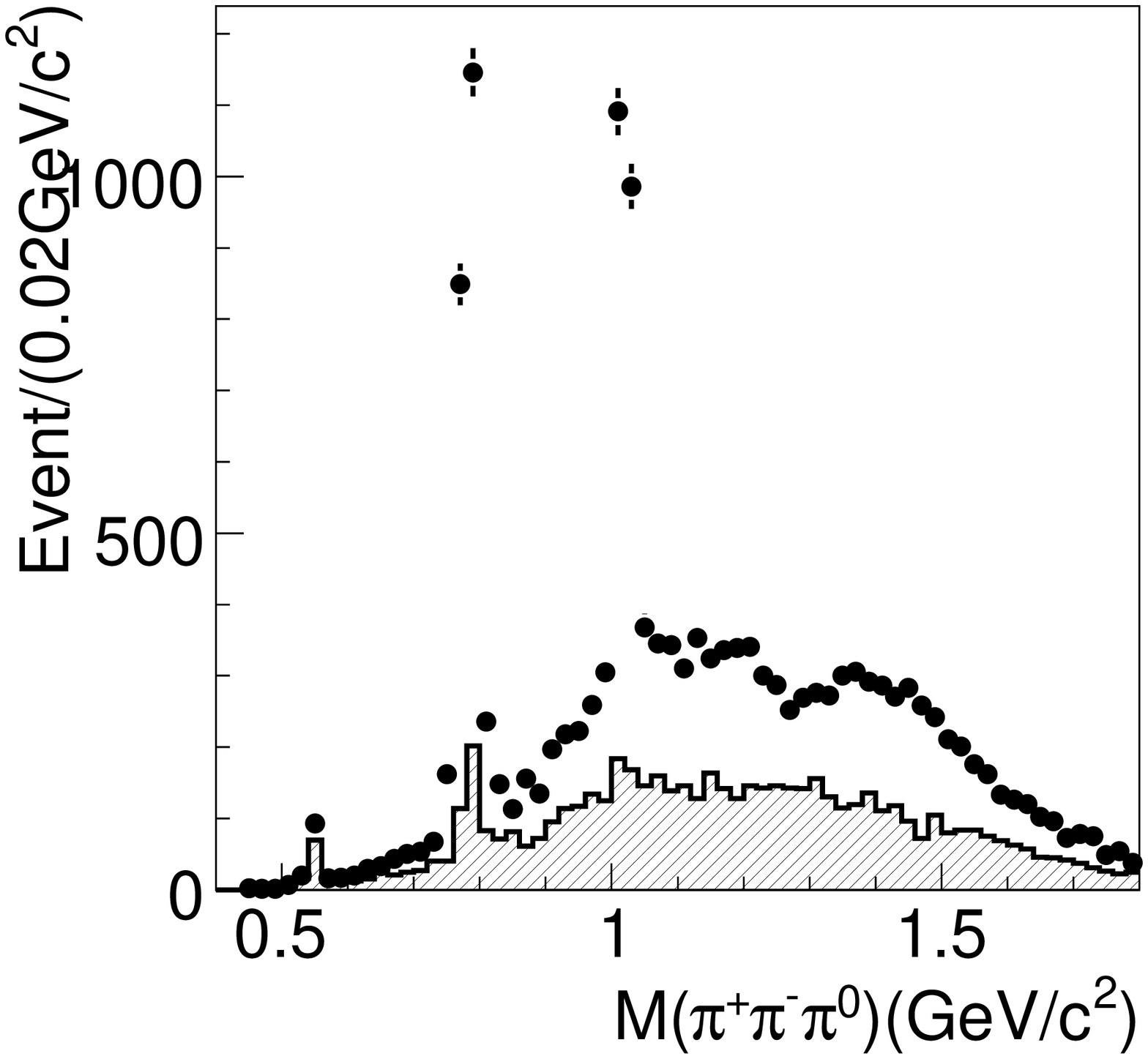}
\put(-25,150){(c)}
}
\vskip -0.5cm
\caption{(a) A scatter plot of $M_{\kk}$ {\it versus} $M_{\ppp}$. The boxes
indicate the signal region labeled as S and sideband regions labeled as
A, B and C (defined in text). (b) The $\kk$ invariant-mass distribution;
the shaded histogram shows the events within the $\omega$ sideband region.
(c) The $\ppp$ invariant mass distribution; the shaded histogram shows
the events within the $\phi$ sideband region.}
\label{scatter}
\end{figure}

\begin{figure}[htbp]
\vskip -0.1 cm \centering
{\includegraphics[width=5.8cm,height=6cm]{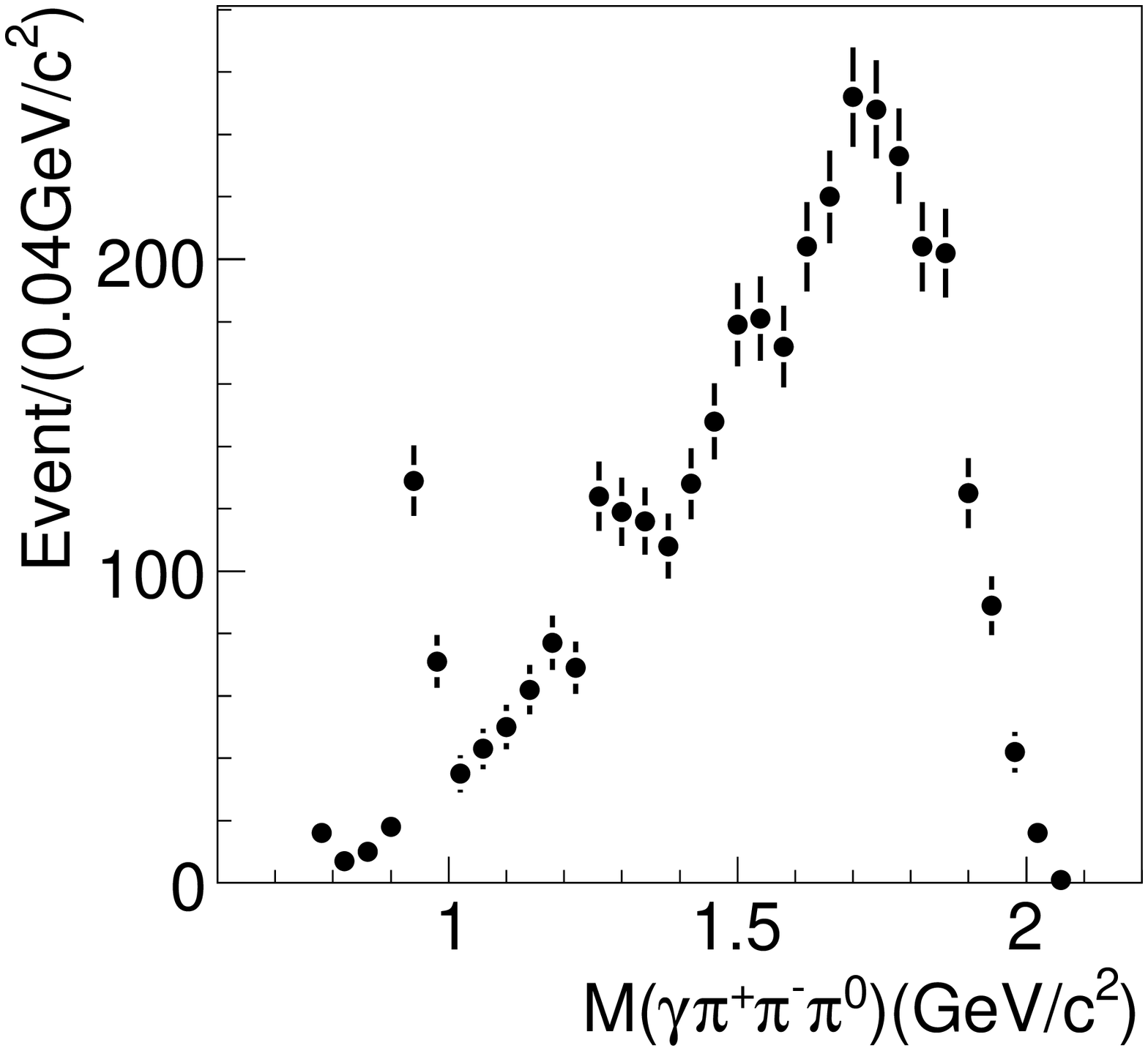}
\put(-25,140){(a)}
\includegraphics[width=5.8cm,height=6cm]{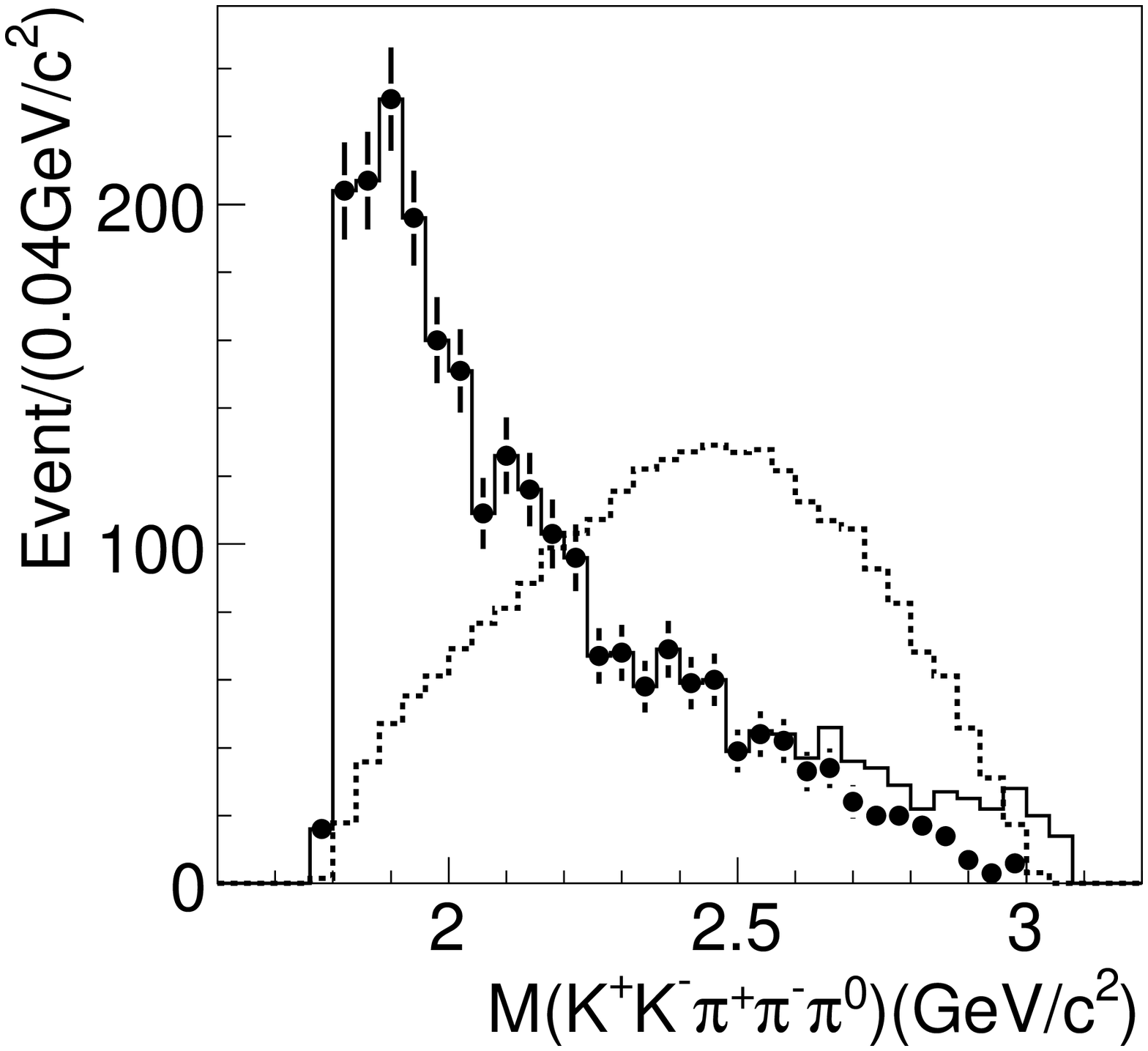}
\put(-25,140){(b)}
\includegraphics[width=5.8cm,height=6cm]{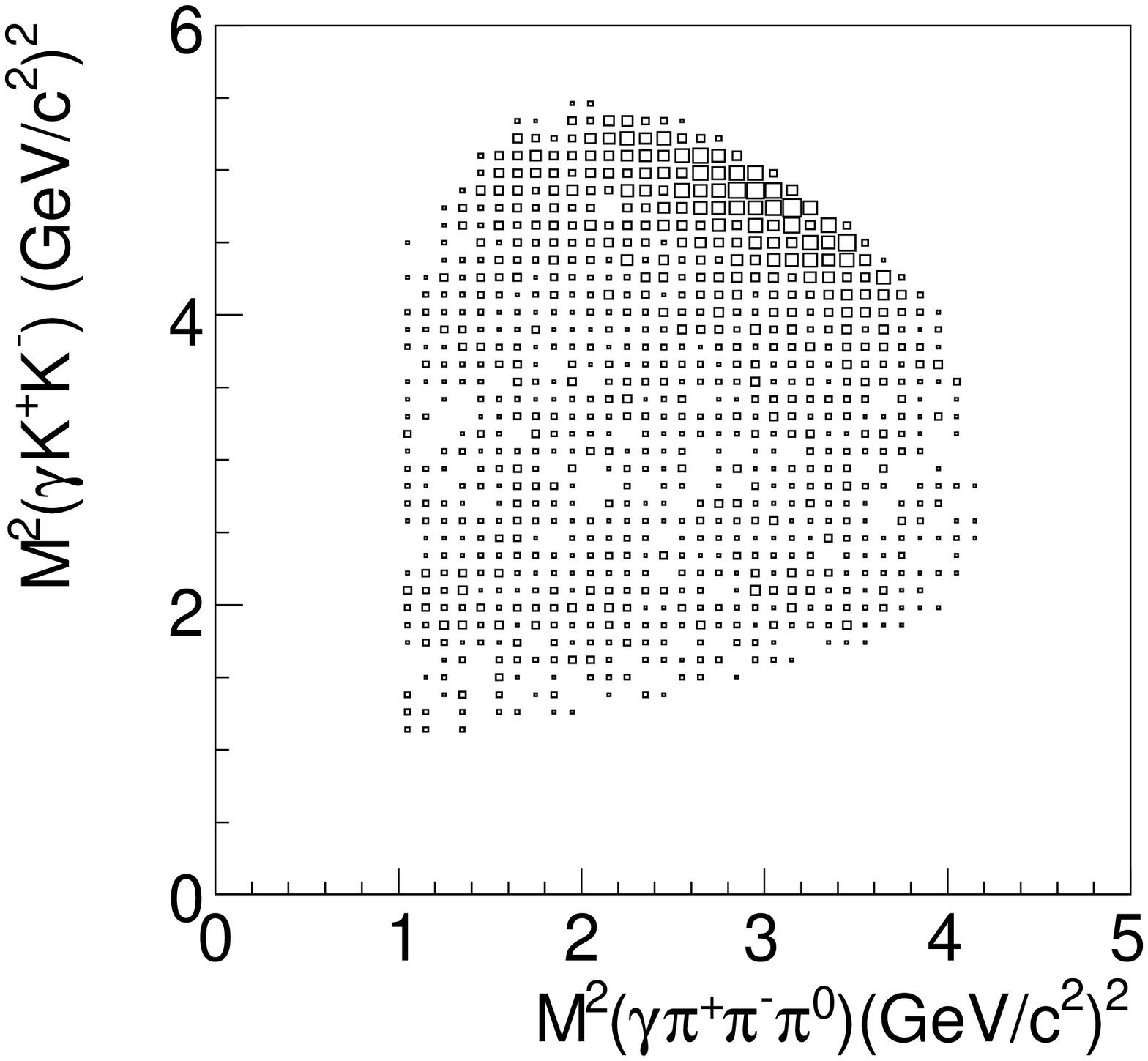}
\put(-25,140){(c)}
}
\vskip -0.5cm
\caption{(a) The $\gamma\ppp$ invariant-mass distribution. (b) The $\kk\ppp$
invariant-mass distribution; the dashed line is the mass distribution
of the phase space MC sample; the solid histogram shows
the mass distribution without the $M(\gamma\ppp$)$>$1.0GeV/$c^{2}$ requirement.
(c) A Dalitz plot of $M^2(\gamma\ppp)$ {\it versus} $M^2(\gamma\kk)$.}
\label{mwf}
\end{figure}
A scatter plot of the $M_{\kk}$ {\it versus} $M_{\ppp}$ invariant masses for events that survive
the above selection criteria is shown in Fig.~\ref{scatter}(a). One cluster
of events populates the $\ff$ region, which arises from the well
known process $\jpsitogff$ (one $\phitoppp$, the other $\phitokk$), and another
cluster of events shows up in the $\wf$ signal region. Since the decays of
$\jpsitowf$ and $\jpsitopwf$ are forbidden by $C$ parity conservation, the
observed events in the $\wf$ region are an unambiguous signal for
the radiative decay process $\jpsitogwf$. The mass window requirements (I)$|M_{\ppp}-M_{\omega}|<$
40MeV/$c^{2}$ (the requirement is determined by optimizing $S$/$\sqrt{S+B}$) and
(II) $|M_{\kk}-M_{\phi}|<15$MeV/$c^{2}$ (the requirement is determined by optimizing $S$/$\sqrt{S+B}$)
are defined for the $\omega$ and $\phi$ signal region, respectively,
while the requirements of (III)
60MeV/$c^{2}<|M_{\ppp}-M_{\omega}|<140$MeV/$c^{2}$ and (IV) 1045MeV/$c^{2}<M_{\kk}<1075$
MeV/$c^{2}$ are defined for the $\omega$ and $\phi$ sideband regions, respectively.
Figure~\ref{scatter}(b) shows the $\kk$ invariant-mass distribution for events
in which the $\ppp$ invariant-mass lies within the $\omega$ signal range (requirement I);
here a $\phi$ signal can clearly be seen. The shaded histogram in Fig.~\ref{scatter}(b)
shows the corresponding distribution for events within the $\omega$ sideband
region (requirement III). A small $\phi$ signal from the $\jpsitogfppp$
background is evident. Figure~\ref{scatter}(c) shows the $\ppp$ invariant-mass
distribution for events with $\kk$ invariant-mass within the $\phi$ signal
range (requirement II). As expected, $\omega$ and $\phi$ signals are clearly seen.
A small $\eta$ signal is also observed; this comes from the decay
chain $\jpsi\to\gamma\eta\kk$($\eta\to\ppp$).
The shaded histogram in Fig.~\ref{scatter}(c) shows the corresponding
distribution for the events within the $\phi$ sideband region (requirement IV).
For events that survive the $\omega$ and $\phi$ requirements
on the $\ppp$ and $\kk$ invariant mass (requirements I and II), respectively, the $\gamma\ppp$
invariant-mass distribution is shown in Fig.~\ref{mwf}(a).
Here an $\etap$ peak is observed; this comes from the decay
chain $\jpsitofetap$($\phitokk$, $\etaptogw$, $\omegatoppp$).
To characterize these events, a large MC sample
of $\jpsitofetap$ is generated with a flat angular distribution.
These have a $\kk\ppp$ invariant mass distribution that is concentrated at masses higher than
2.5GeV/$c^{2}$ and have no impact on the $\omega\phi$ mass threshold region of interest.
A further requirement $M(\gamma\ppp$)$>$1.0GeV/$c^{2}$ (requirement V) is imposed to remove
background from $\jpsitofetap$.
Figure~\ref{mwf}(b) shows the invariant mass of $\kk\ppp$ for events with requirements I, II and V
applied, where a peaking structure near the
$\wf$ invariant-mass threshold is observed. The solid histogram in the figure shows
the $\kk\ppp$ invariant-mass distribution without requirement V.
The invariant mass distribution
is very different from a pure phase-space distribution from MC (dashed
histogram, arbitrarily scaled). The threshold structure shows up as a diagonal band
along the upper right-hand edge of the Dalitz plot in Fig.~\ref{mwf}(c).

\begin{figure}[htbp]
\vskip -0.1 cm \centering
{\includegraphics[width=6cm,height=6cm]{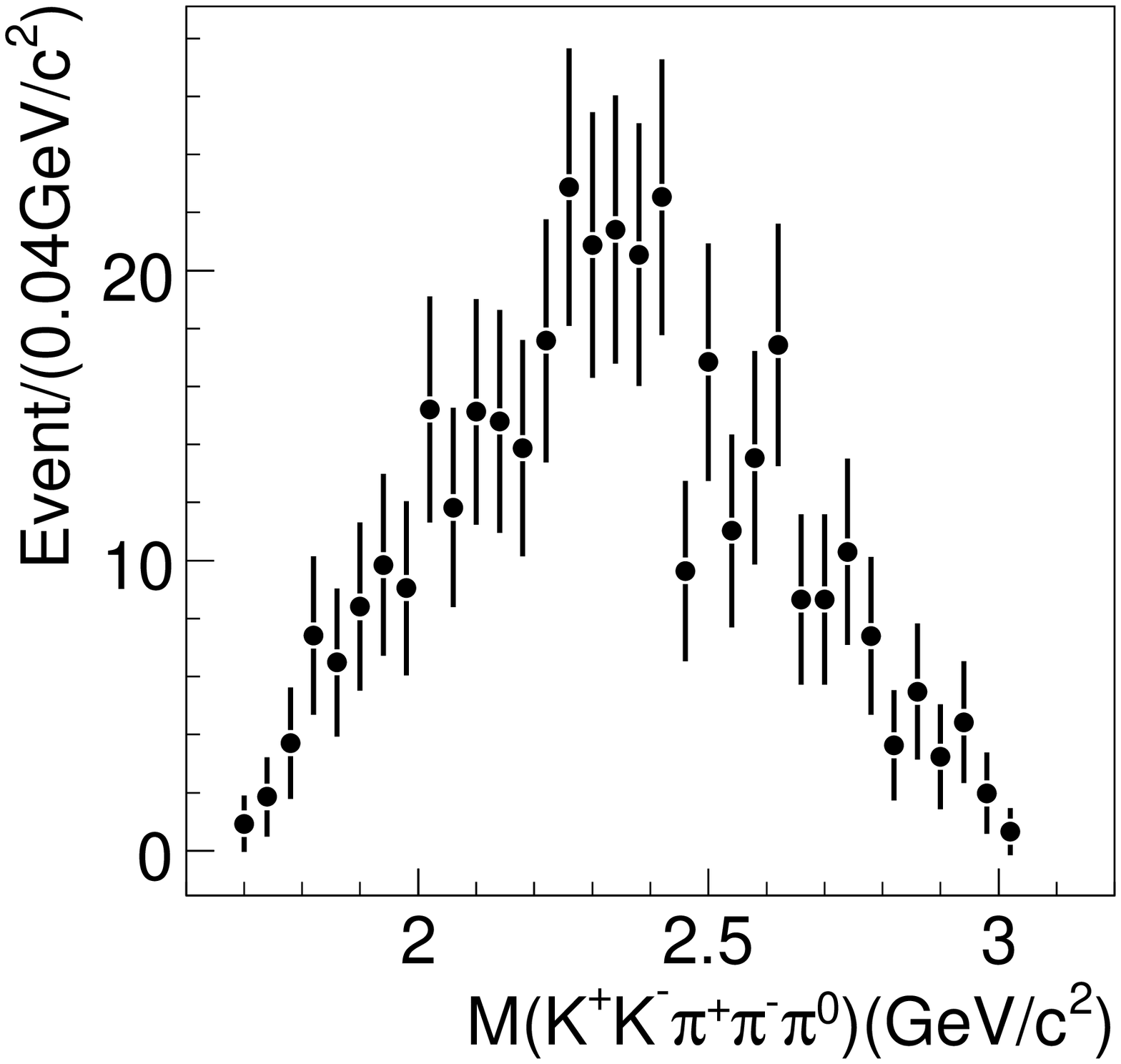}
\put(-25,140){(a)}
\includegraphics[width=6cm,height=6cm]{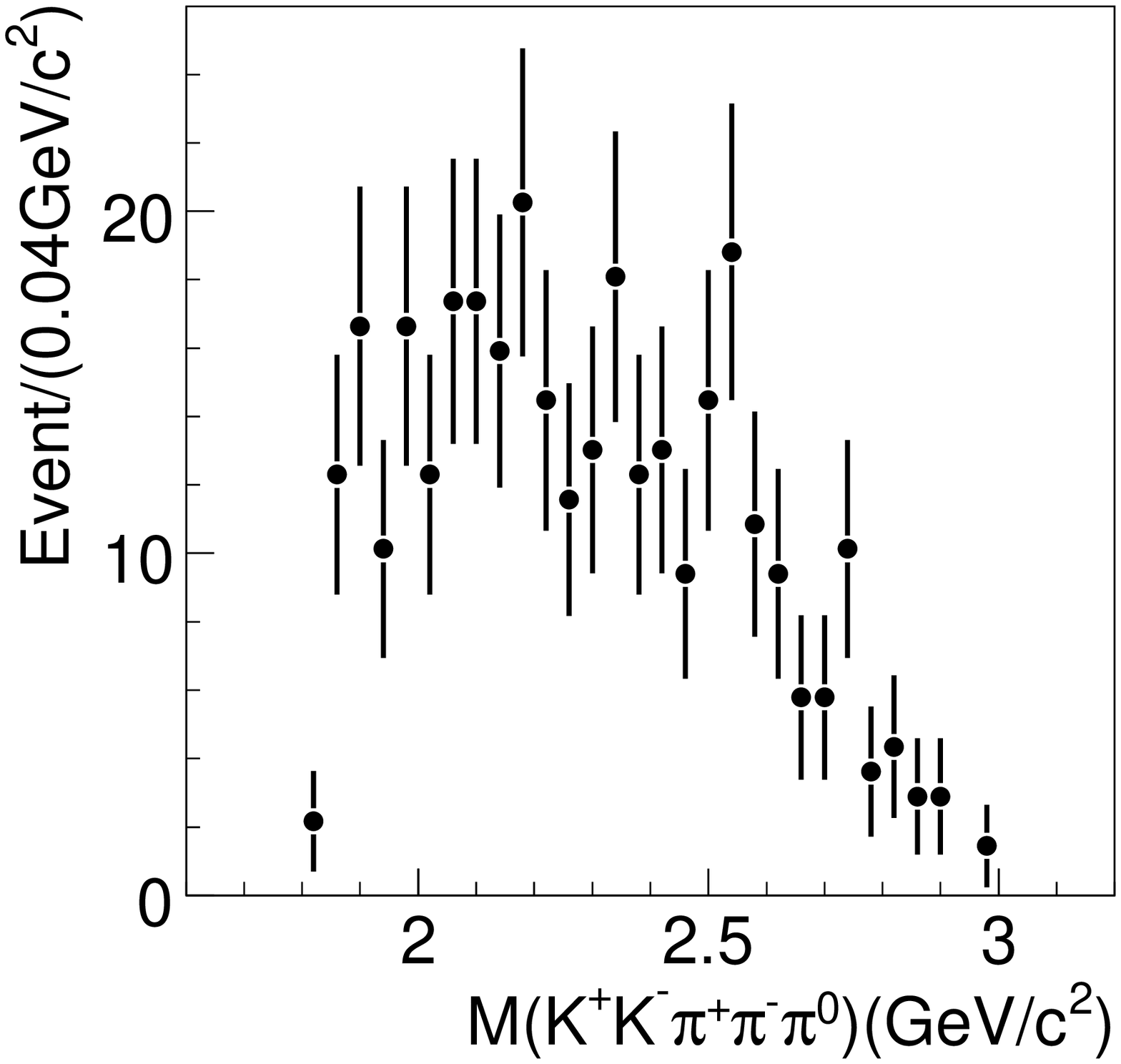}
\put(-25,140){(b)}\\
\includegraphics[width=6cm,height=6cm]{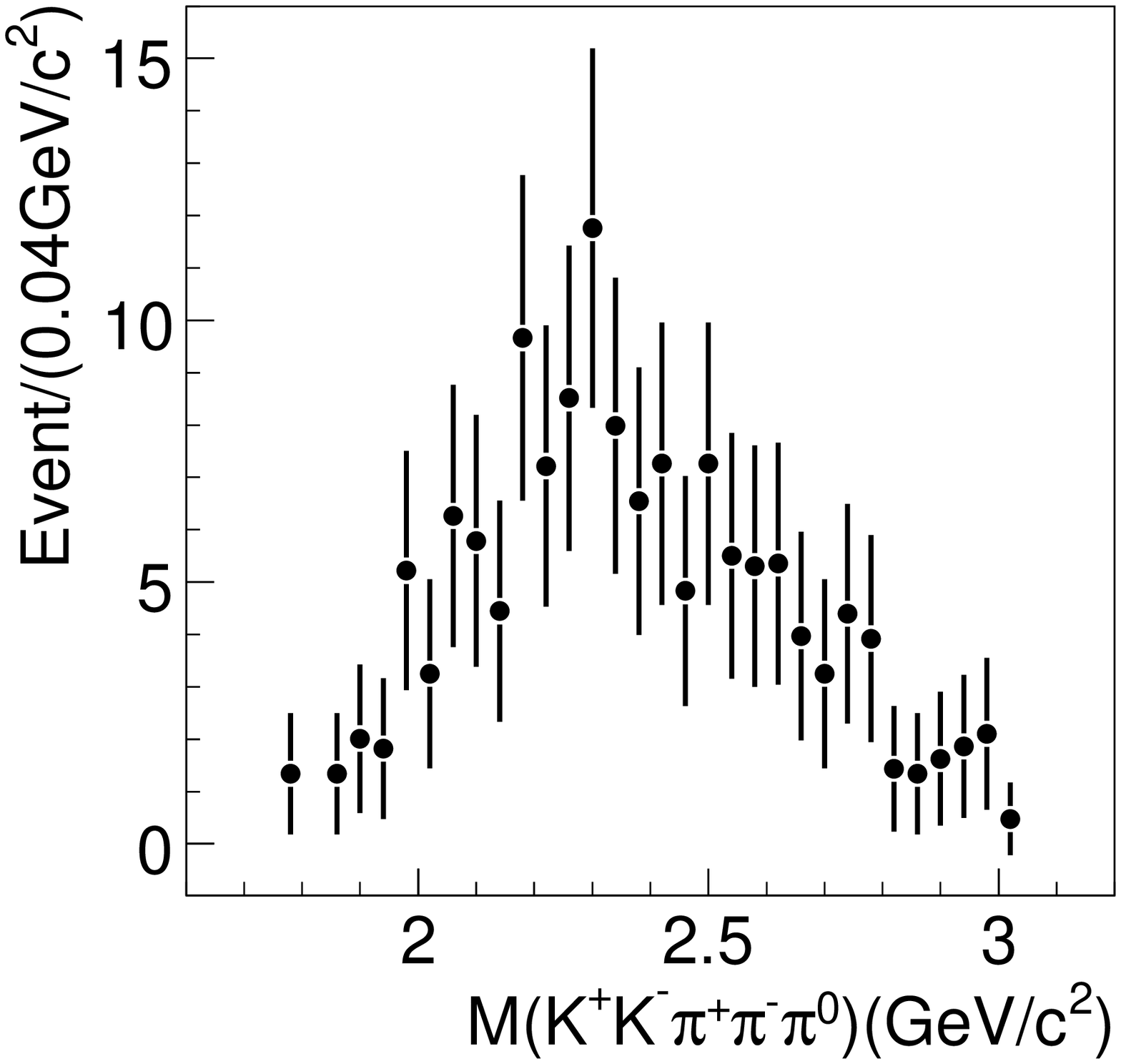}
\put(-25,140){(c)}
\includegraphics[width=6cm,height=6cm]{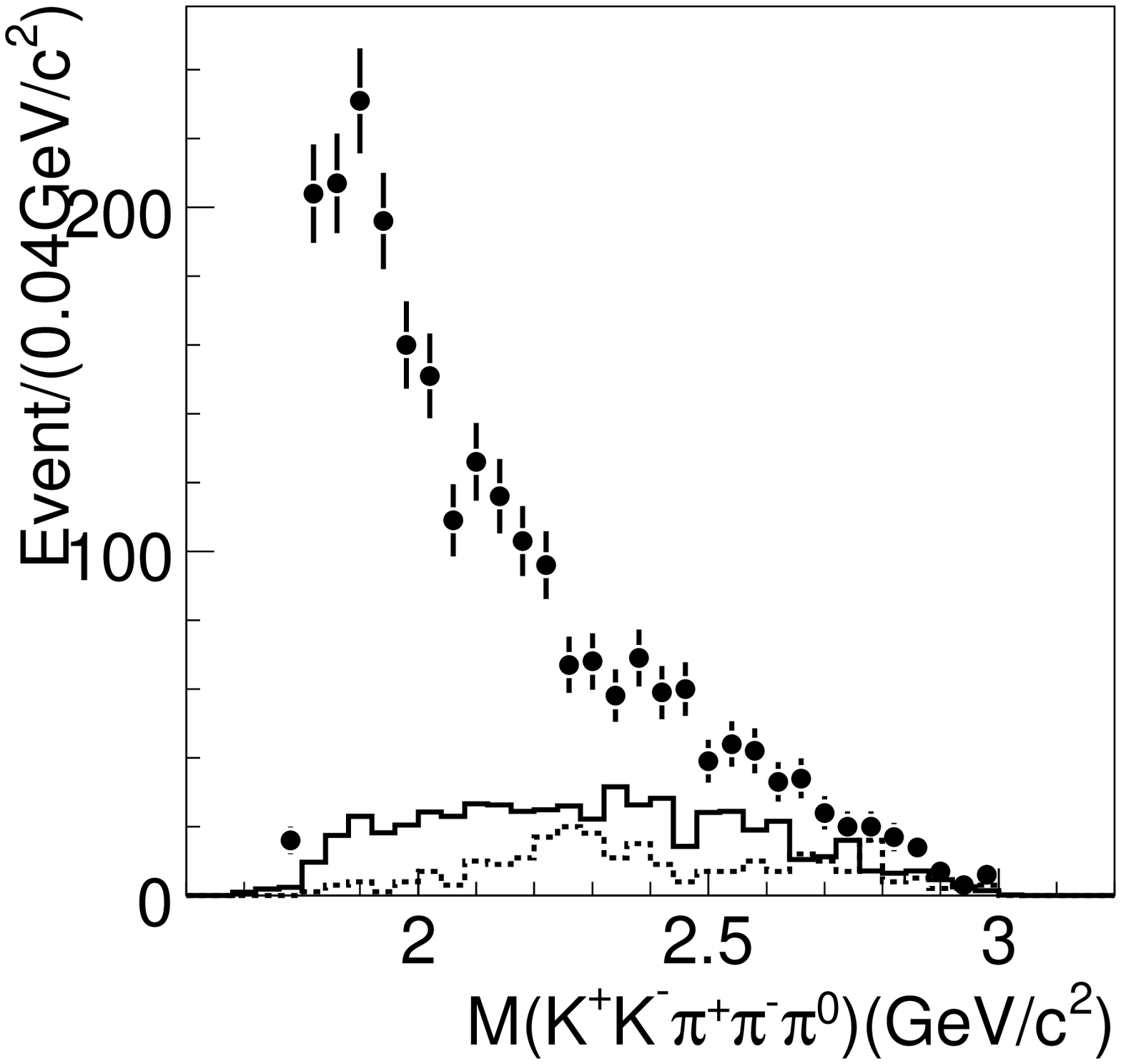}
\put(-25,140){(d)}
}
\caption{
The $\kk\ppp$ invariant-mass distribution for (a) the events in the
$\omega$ sideband region (Box A in Fig.~\ref{scatter}(a));
(b) the events in the $\phi$ sideband region (Box B in Fig.~\ref{scatter}(a));
(c) the events in the corner region (Box C in Fig.~\ref{scatter}(a));
(d) the events in the $\wf$ signal
region; the solid histogram is the background distribution estimated
from the sideband events, the dashed histogram is that obtained from
inclusive $\jpsi$ MC samples.
}
\label{sideband}
\end{figure}

The observed $\omega\phi$ mass-threshold enhancement is
similar to that observed by the BESII experiment~\cite{Ablikim:2006}. To ensure that the
enhancement is not due to some background process, detailed studies of potential
background sources have been performed using both data and MC.
Non-$\omega$ and non-$\phi$ backgrounds are studied using
$\omega$ and $\phi$ sideband data. Figure~\ref{sideband}(a) and ~\ref{sideband}(b) show
the $\kk\ppp$ invariant mass for events in the $\omega$
sideband region (labeled as Box A in Fig.~\ref{scatter}(a)) and the
$\phi$ sideband region (labeled as Box B in Fig.~\ref{scatter}(a));
these are used to determine the non-$\omega$ and non-$\phi$ background
contamination in the signal regions. Figure~\ref{sideband}(c)
shows the same distribution for events in the corner region
(labeled as Box C in Fig.~\ref{scatter}(a)), for which both the $\kk$ and the $\ppp$
invariant masses are in the $\phi$ and $\omega$ sidebands;
these are used to estimate the non-$\phi$ non-$\omega$ background.
The background contamination in the signal region is estimated to be the sum of the
Fig.~\ref{sideband}(a) and Fig.~\ref{sideband}(b) sideband distributions with
the Fig.~\ref{sideband}(c) distribution subtracted to account for double counting
of non-$\phi$ non-$\omega$ background in Fig.~\ref{sideband}(a) and
Fig.~\ref{sideband}(b). Phase-space-MC-determined normalization
factors are applied that account for differences in the sizes of the selected regions and the
difference in the available phase space in the signal and sideband regions.
The background contamination in the signal region determined in this way
is shown as a solid histogram in
Fig.~\ref{sideband}(d). The shape of the estimated background is very different
from that of data in the signal region, and no evidence of an enhancement near the
$\wf$ mass threshold is observed from the non-$\omega$ and non-$\phi$ background
events in the data.

An inclusive MC sample of 225M $\jpsi$ events generated according to the Lund-Charm
model~\cite{lundcharm} and the PDG decay tables is also used to
study the potential backgrounds. The dashed histogram in
Fig.~\ref{sideband}(d) shows the $\kk\ppp$ invariant-mass distribution for
the selected inclusive $\jpsi$ MC events, where no peaking background at the $\wf$ invariant-mass
threshold is observed. Exclusive background MC samples
of $\jpsi$ decays that have similar final states are generated to further
investigate possible background sources. The main backgrounds come from
$\jpsi\to\omega K^*K, K^*\to K\piz$ and $\jpsi\to\omega f_1(1420), f_1(1420)\to\kk\pi^0$
events. For these, the $\kk\ppp$ invariant mass distribution peaks at high masses,
and none of them channels produce peaking structures at the $\of$ mass threshold.

\section{Partial Wave Analysis}
A PWA was performed on the selected $\jpsitogwf$
candidate events to study the properties of the $\of$ mass threshold enhancement.
In the PWA, we assume the enhancement is due to the presence of a resonance, denoted as $X$, and the
decay processes are described with sequential 2-body or 3-body decays:
$\jpsi\to\gamma X, X\to\wf$, $\omegatoppp$ and $\phitokk$. The amplitudes
of the 2-body or 3-body decays are constructed with a covariant tensor
amplitude method~\cite{tensor}. The intermediate structure $X$ is parameterized
with the Breit-Wigner propagator
\beq
BW =1/(M^{2}-s-iM\Gamma)
\eeq
with constant width, where $s$ is the $\wf$ invariant mass-squared, and $M$ \&
$\Gamma$ are the resonance mass and width, respectively. The amplitude for the sequential
decay process is the product of all decay amplitudes together with the
Breit-Wigner propagator. The total differential cross section $d\sigma/d\Phi$
for the process is the square of the linear sum of all possible
partial wave amplitudes:
\beq
 \frac{d\sigma}{d\Phi}=|\sum A(J^{PC}) |^2,
 \label{equ1}
\eeq
where $A(J^{PC})$ is the total amplitude for all possible resonances with
given $J^{PC}$.

The relative magnitudes and phases of the states are determined by an
unbinned maximum likelihood fit of the measured cross section $d\sigma/d\Phi$.
The basis of likelihood fitting is
the calculation of the probability that a hypothesized probability distribution
function can produce the data set under consideration. The probability
to observe the event characterized by the measurement $\xi_i$ is the
differential cross section normalized to unity:
\beq
P(\xi_i)=\frac{\omega(\xi_i)\epsilon(\xi_i)}{\int d\xi_i\omega(\xi_i)\epsilon(\xi_i)},
\eeq
where $\omega(\xi_i)\equiv(\frac{d\sigma}{d\Phi})_i$ and $\epsilon(\xi_i)$
is the detection efficiency. The joint probability density for observing the
$N$ events in the data sample is:
\beq
 \mathcal{L} = \prod^{N}_{i=1} P(\xi_{i}) = \prod^{N}_{i=1}
\frac{\omega(\xi_i)\epsilon(\xi_i)}{\int d\xi_i\omega(\xi_i)\epsilon(\xi_i)}.
\eeq
FUMILI~\cite{FUMILI} is used to optimize the fit parameters in order to achieve
the maximum likelihood value. Technically, rather than maximizing $\mathcal{L}$,
$\mathcal{S}$ = -ln$\mathcal{L}$ is minimized, i.e.,
\beq
\mathcal{S} = -ln\mathcal{L} = -\sum^{N}_{i=1} ln(\frac{\omega(\xi_i)}{\int d\xi_i\omega(\xi_i)\epsilon(\xi_i)})
            - \sum^{N}_{i=1}ln\epsilon(\xi_i).
\eeq
In practice, the normalized integral $\int d\xi_i\omega(\xi_i)\epsilon(\xi_i)$
is evaluated using the $\jpsitogwf$ phase space MC sample. For a given data set, the second
term is a constant and has no impact on the relative changes of the $\mathcal{S}$
value. The details of the PWA fit process are described in Ref.~\cite{goodness}.
In the minimization procedure, a change in log likelihood of 0.5 represents
a one standard deviation effect for the one-parameter case and is used to
evaluate statistical errors.

Conservation of $\JPC$ , in the $\jpsi\to\gamma X, X\to\wf$ process in the case of
a pseudoscalar intermediate resonance $X$, allows only
$\mathcal{P}$ wave contributions in both the radiative decay
$\jpsi\to\gamma X$ and the hadronic decay $X\to\wf$. For the production
of a $0^{++}$, $1^{++}$ or $2^{++}$ resonance, both $\mathcal{S}$
and $\mathcal{D}$ waves are possible for both the radiative and hadronic
decays, but only the $\mathcal{S}$ wave contribution is considered in
the fit, since the $\mathcal{D}$ wave can be expected to be
highly suppressed near the mass threshold.
Intermediate $X$ structures with $\JPC=2^{-+}$ or higher
spin are not considered in the analysis. To investigate the $\JPC$ of
the $X(1810)$, we tried different $\JPC$ assignments
in the fit, and the assignment with the best log likelihood
value is identified as the $\JPC$ of the $X(1810)$. Some known mesons,
e.g. $f_{2}$(1950) or $f_0(2020)$, with a mass above the $\wf$ invariant-mass
threshold, are expected to decay to $\wf$ final states. All possible
mesons listed in the PDG tables are included in the fit.
To consider the contribution from phase space, $\jpsitogwf$
without an intermediate state $X$ is also included in the fit
with an amplitude modeled by the same sequential process and a very broad
intermediate state, i.e.,
$M$ = 2500MeV and $\Gamma$ = 5000MeV. In the PWA fit, the phase space is
assigned to a given $\JPC$, which is determined by the optimization of
the likelihood fit. The background event contribution to the log likelihood
value is estimated from the weighted events in the sideband region, and
subtracted in the fit.

In the PWA fit, different $J^{PC}$ combinations of the $X(1810)$ structure
and the phase-space contribution, as well as different combinations of additional
mesons listed in the PDG tables, are tried. The mass and width of the
$X(1810)$ are determined by a scan of the maximum log likelihood value,
while the mass and width of the additional mesons are fixed with their PDG values.
The statistical significance of the state is determined by the changes
of the maximum log likelihood value and of the number of degrees of freedom ($\Delta ndf$)
in the PWA fits with or without the state included. Only states with statistical
significance larger than 5$\sigma$ are included in the best solution.

\begin{figure}[htbp]
\vskip -0.1 cm \centering
{\includegraphics[width=5.8cm,height=6cm]{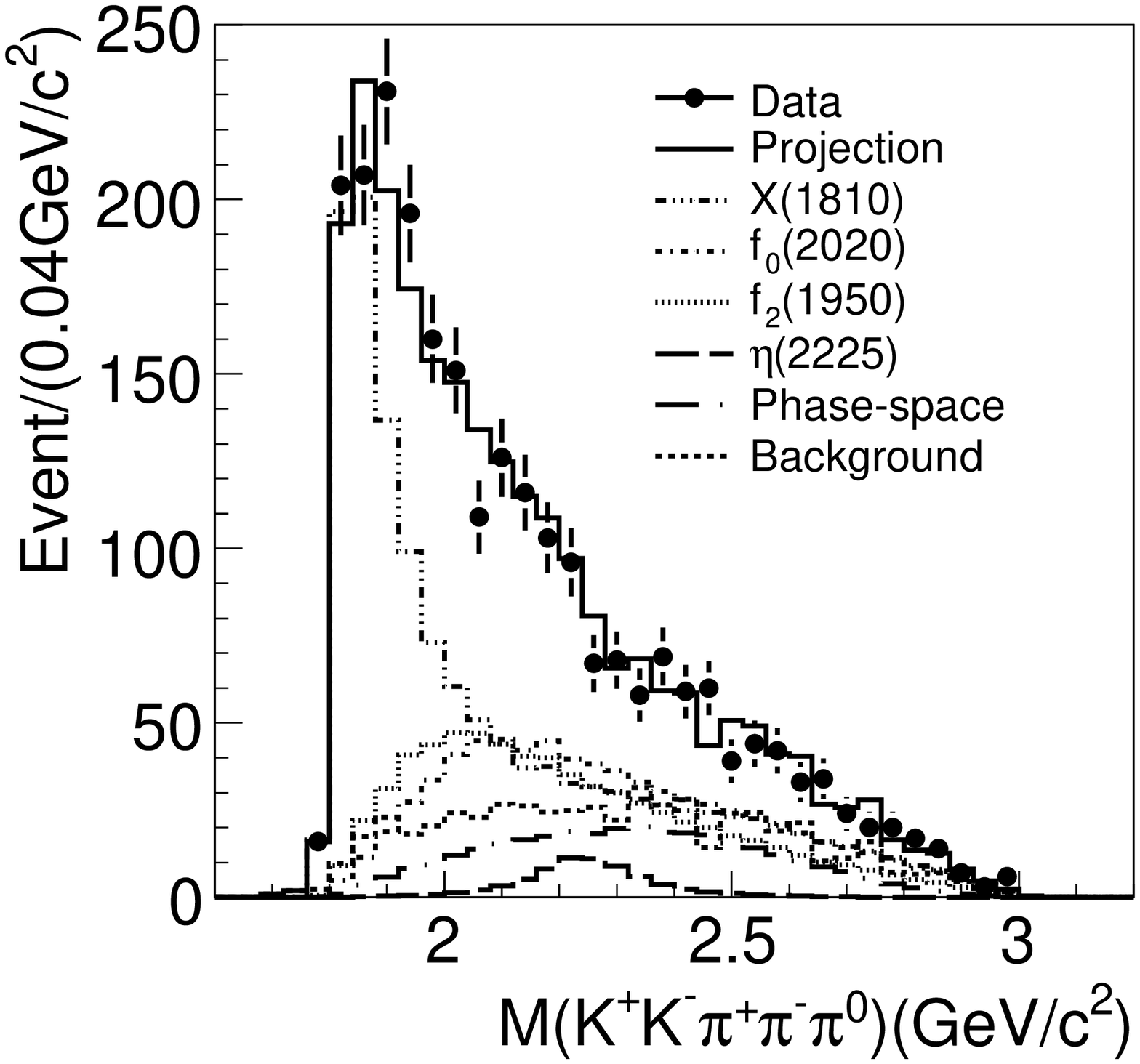}
\put(-25,140){(a)}
\includegraphics[width=5.8cm,height=6cm]{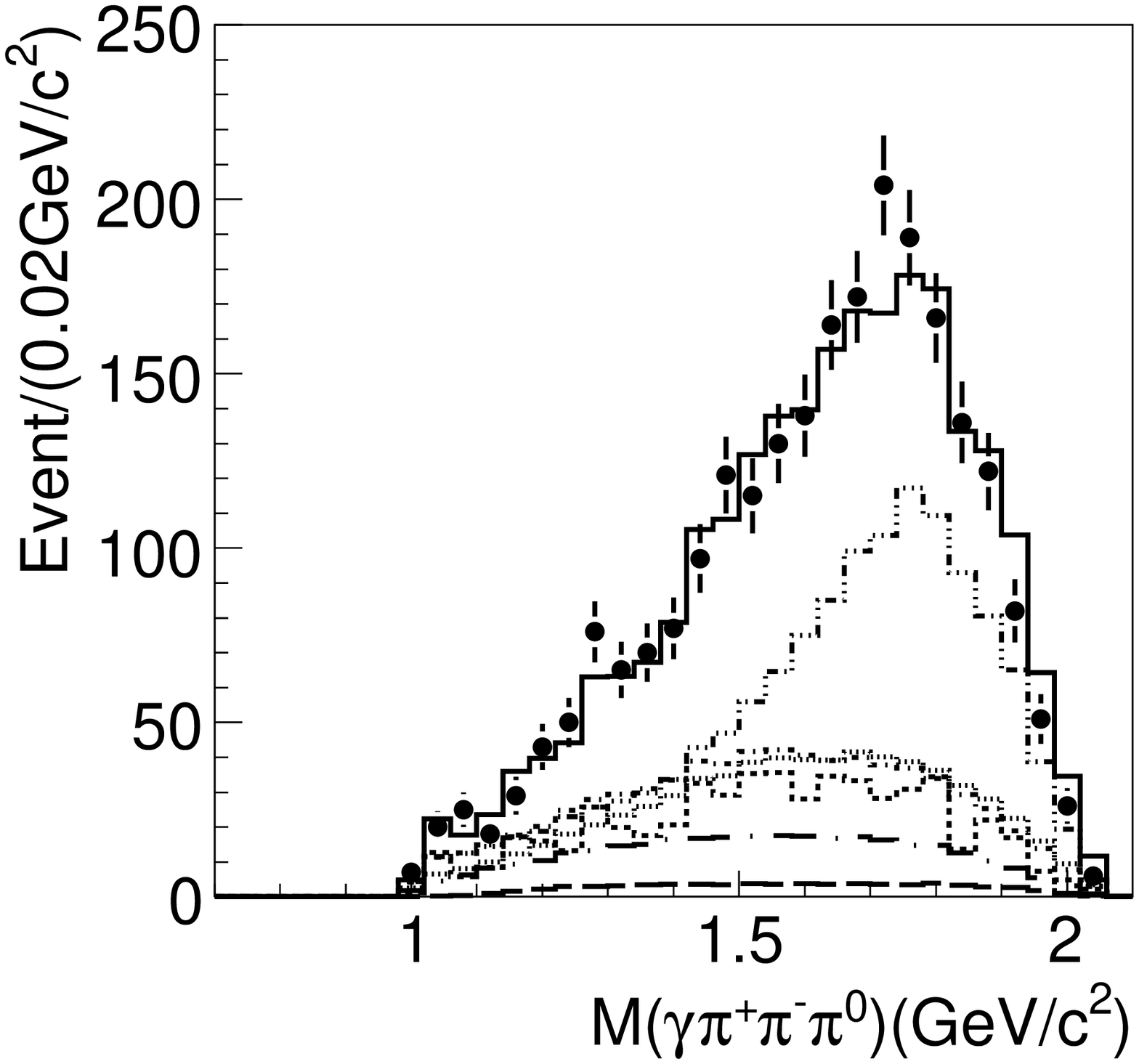}
\put(-25,140){(b)}
\includegraphics[width=5.8cm,height=6cm]{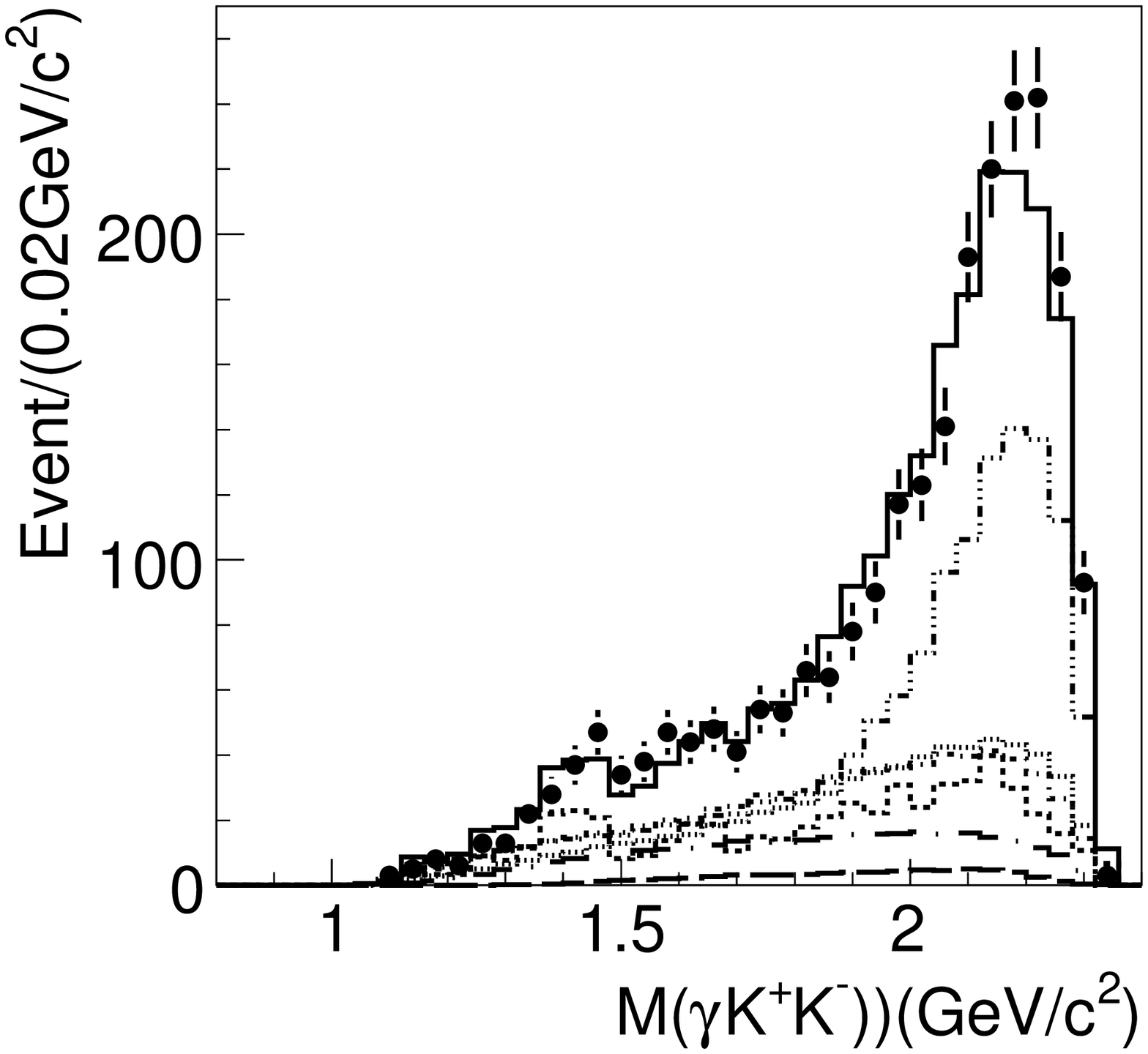}
\put(-25,140){(c)}\\
\includegraphics[width=5.8cm,height=6cm]{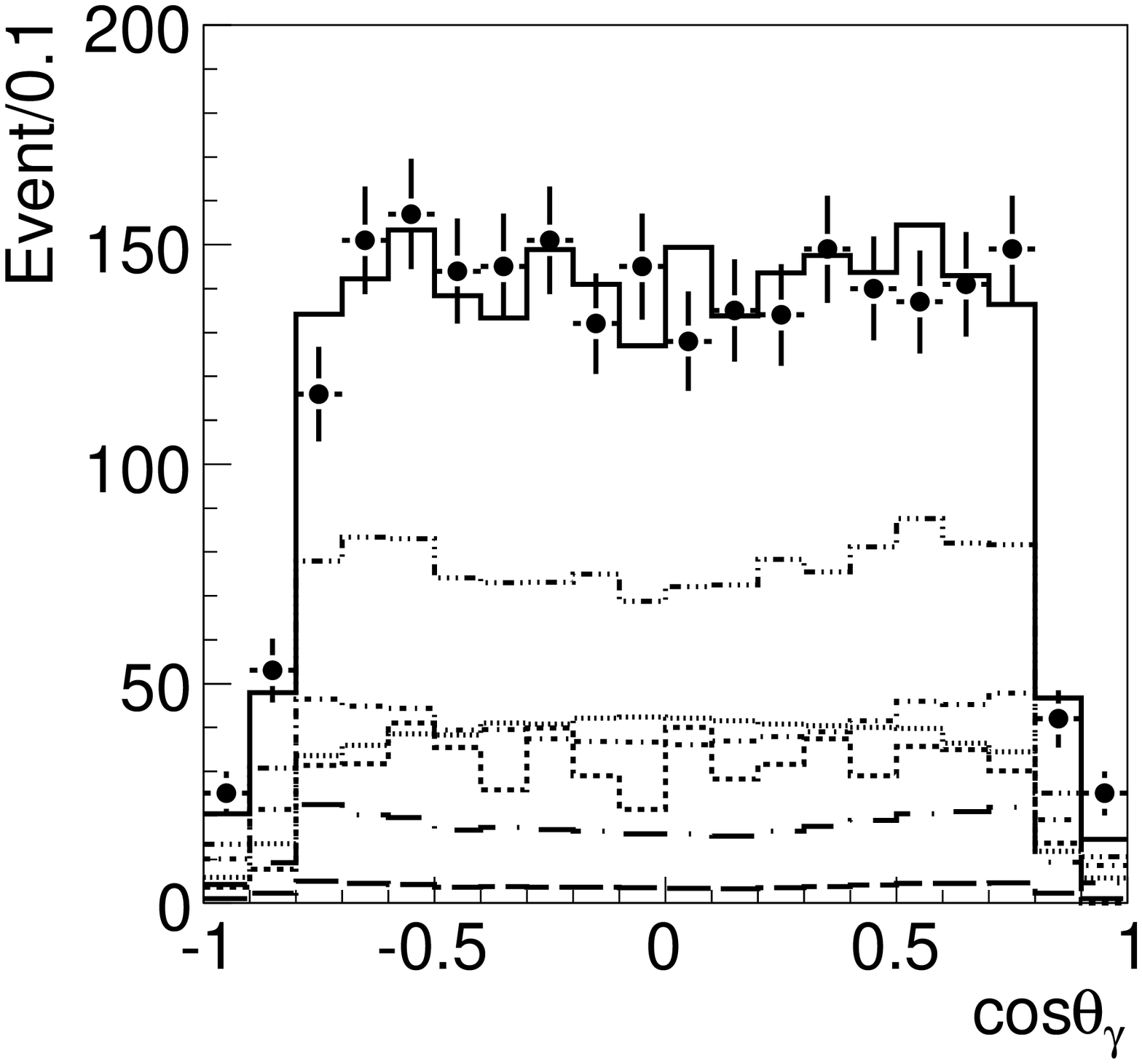}
\put(-25,140){(d)}
\includegraphics[width=5.8cm,height=6cm]{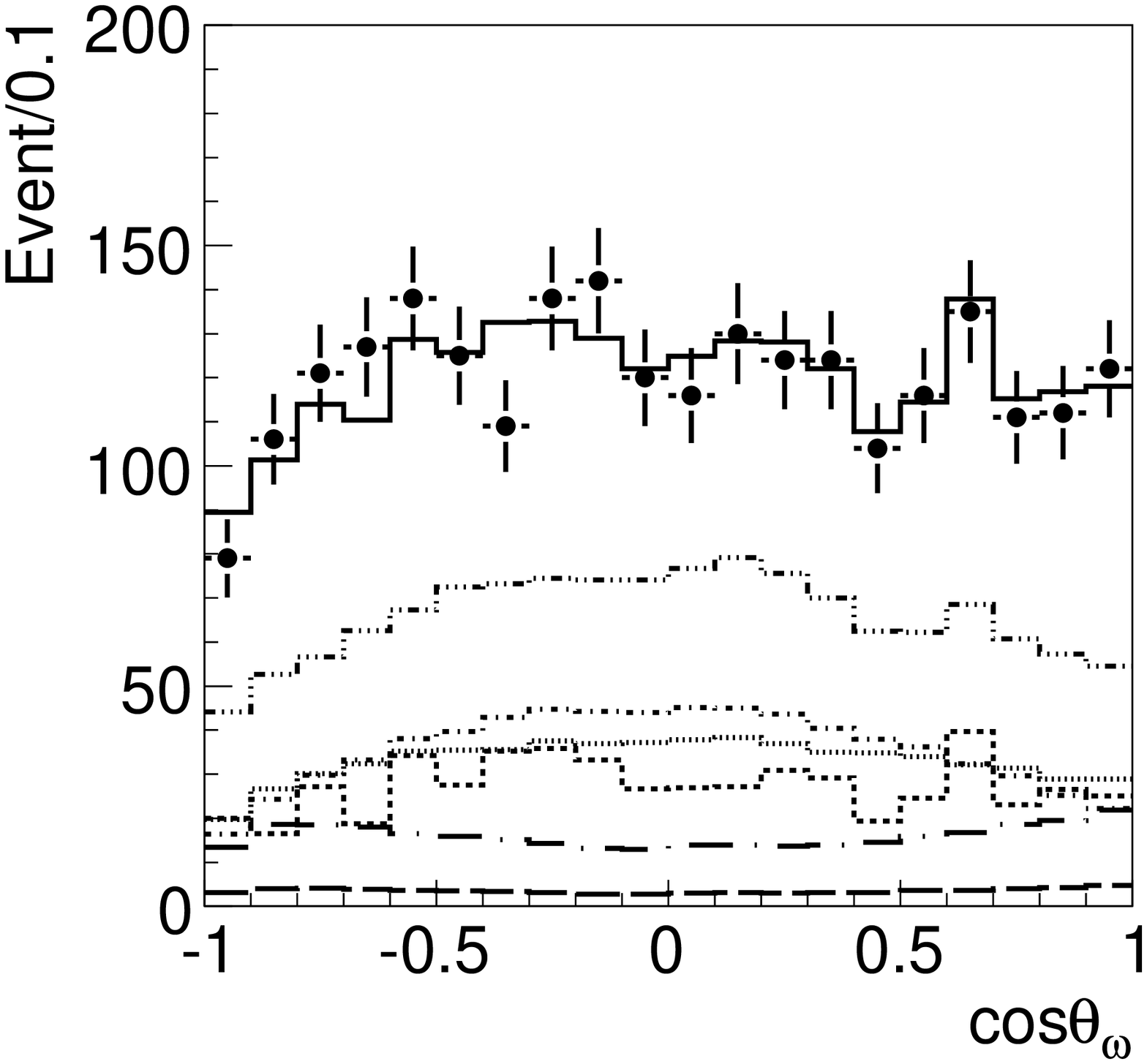}
\put(-25,140){(e)}
\includegraphics[width=5.8cm,height=6cm]{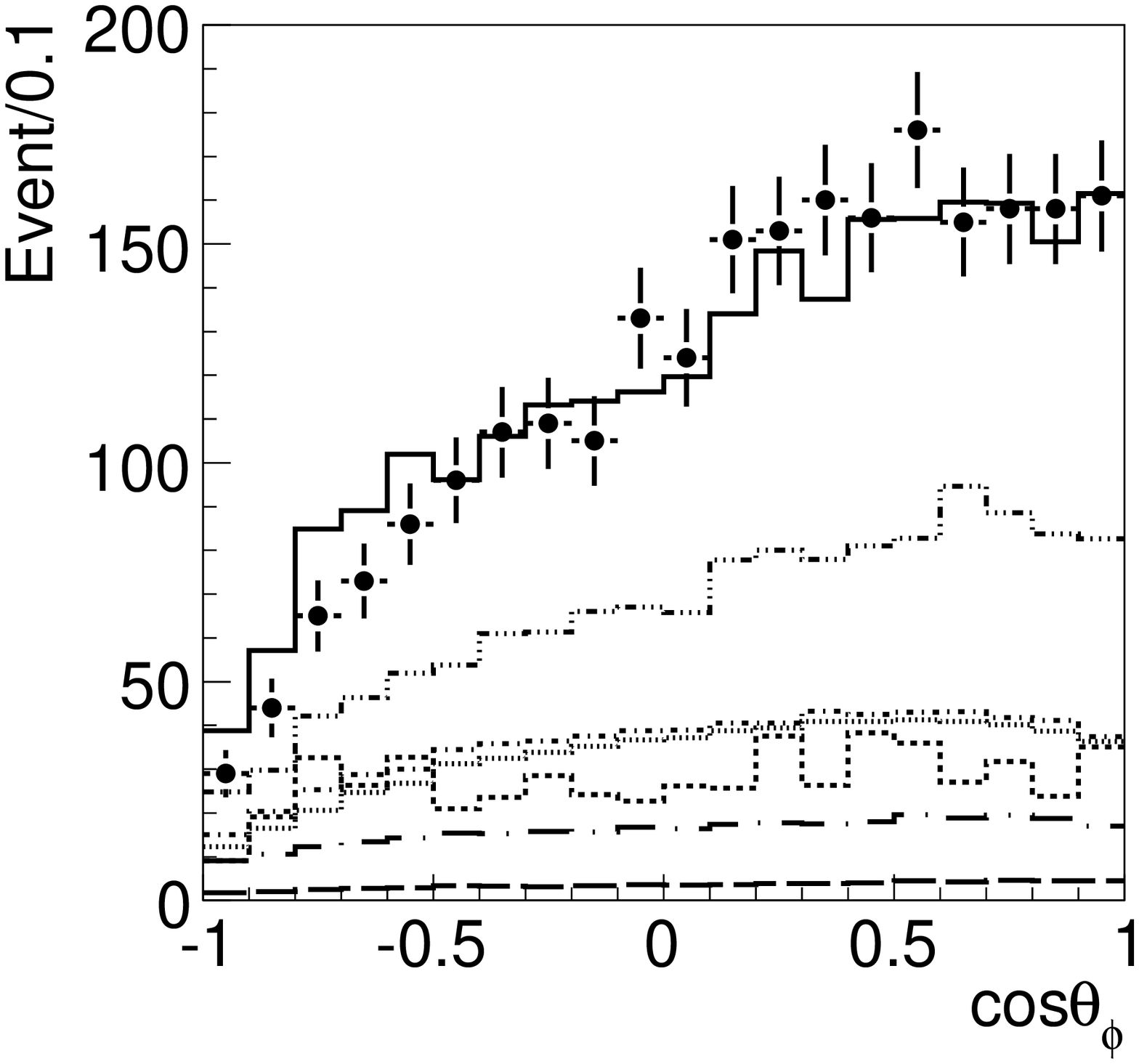}
\put(-25,140){(f)}\\
\includegraphics[width=5.8cm,height=6cm]{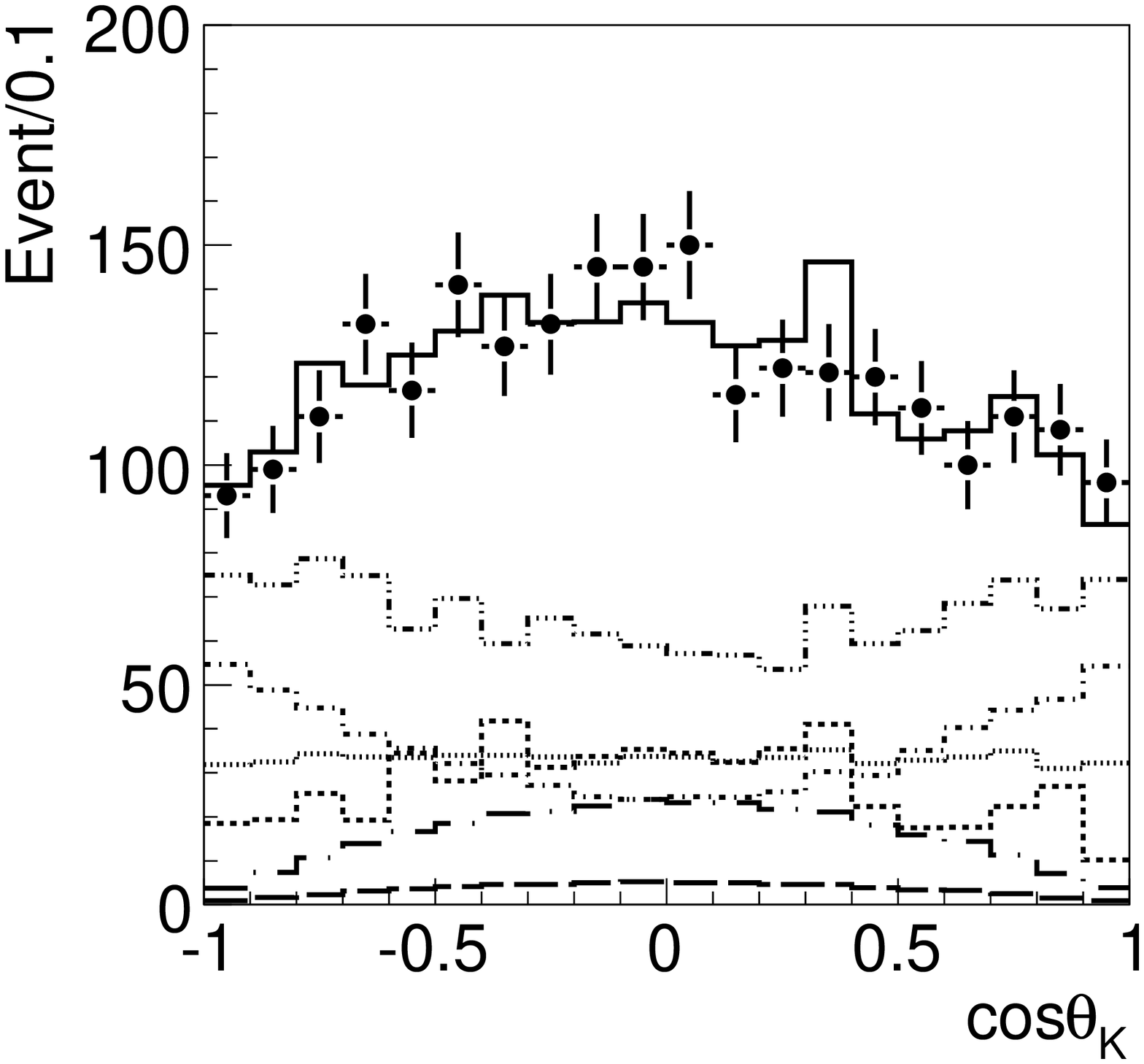}
\put(-25,140){(g)}
\includegraphics[width=5.8cm,height=6cm]{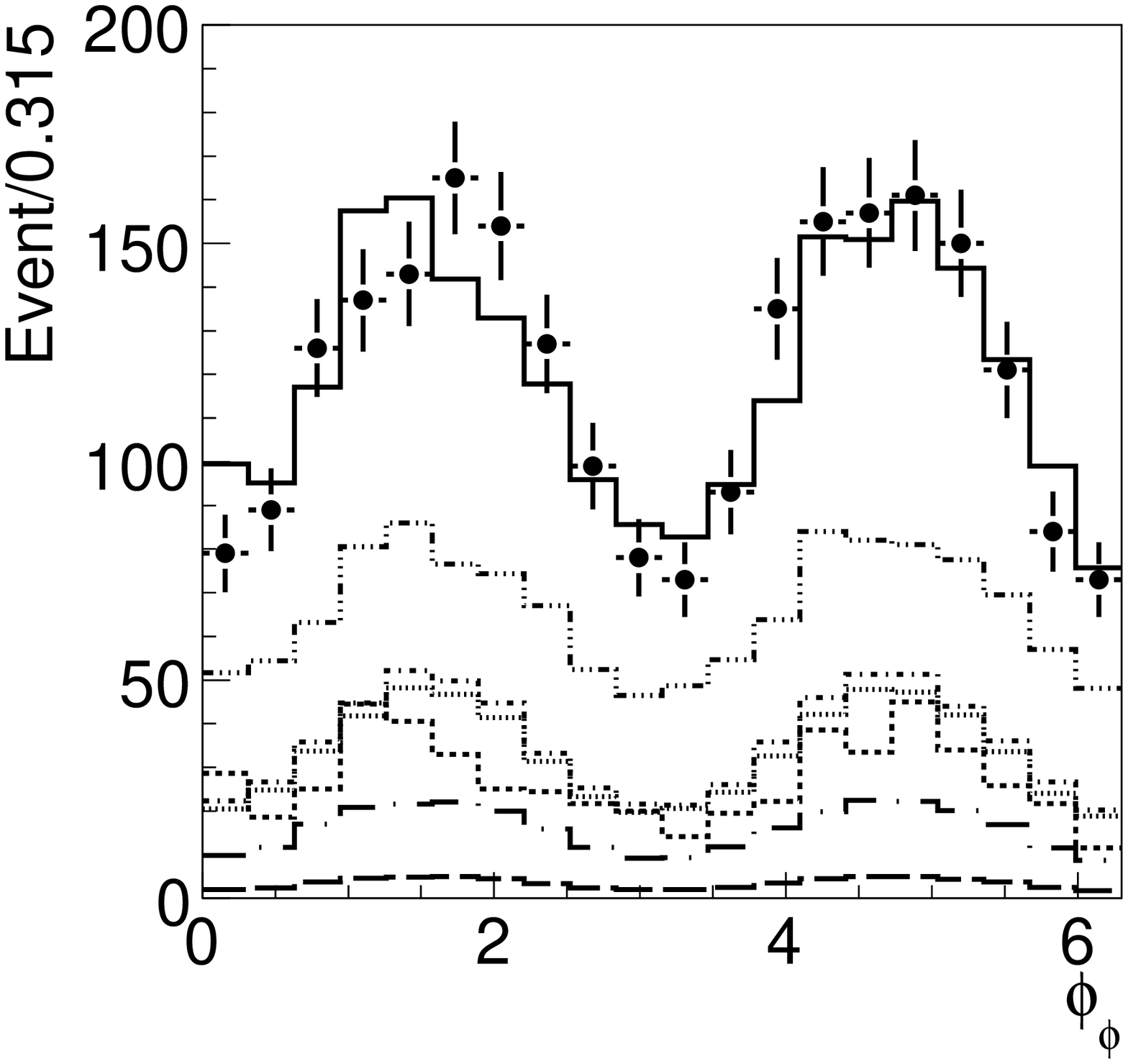}
\put(-25,140){(h)}
\includegraphics[width=5.8cm,height=6cm]{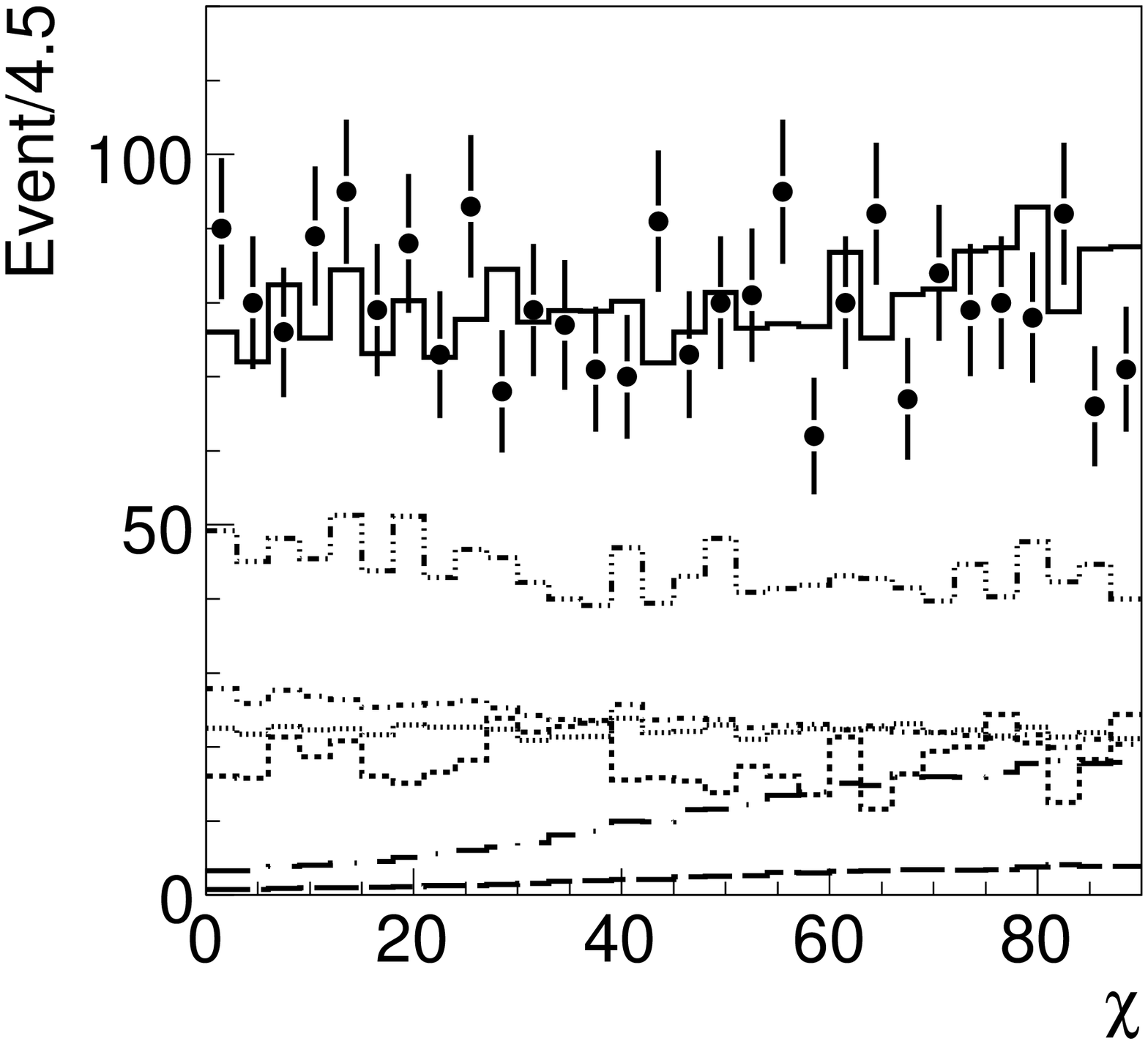}
\put(-25,140){(i)}
}
\vskip -0.5cm
\caption{Comparisons between data and PWA fit projections:
(a) The $\kk\ppp$ invariant-mass distribution;
(b) the $\gamma\ppp$ invariant-mass distribution;
(c) the $\gamma\kk$ invariant-mass distribution;
(d) the polar angle of the radiative photon ($\theta_{\gamma}$);
(e) the polar angle of the normal
to the $\omega$ decay plane in the $\omega$ system ($\theta_{\omega}$);
(f) the polar angle of the $\phi$ in the $\of$ rest system ($\theta_{\phi}$);
(g) the polar angle of the kaon in the $\phi$ rest system ($\theta_K$);
(h) the azimuthal angle of the $\phi$ in the $\of$ system;
(i) the distribution of $\chi$ which is the angle between azimuthal angles
of the normal to the $\omega$ decay plane and the momentum of a kaon from
$\phi$ decay in the $\of$ system.
}
\label{pwares}
\end{figure}

Finally, together with the contributions of the $X(1810)$ and phase-space, additional
0$^{++}$, 2$^{++}$, and 0$^{-+}$ components are found ($>5\sigma$)
in the best solution of the PWA fit.
In the following, the masses and widths of the 0$^{++}$, 2$^{++}$
and 0$^{-+}$ components are assigned to be those of $f_{0}$(2020), $f_{2}$(1950) and
$\eta$(2225), respectively, since the fit with these has
the best log likelihood value.
Various PWA fits with different 0$^{++}$, 2$^{++}$ and 0$^{-+}$ components were also performed.
The results for the $X(1810)$ are robust,
while the fit is not very sensitive to the masses and widths of
the 0$^{++}$, 2$^{++}$ and 0$^{-+}$ components. The log likelihood values changes are rather small
when the $f_{0}$(2020), $f_{2}$(1950) and $\eta$(2225) are replaced by other resonances with
the same $\JPC$ and similar masses. The details are shown below.
The $\JPC=0^{++}$ assignment for the $X(1810)$ has by far
the highest log likelihood value among the different $\JPC$ hypotheses.
The minus log likelihood value ($S$) for a $\JPC=0^{++}$ assignment to the $X(1810)$ is
227 below that of the second lowest value (obtained for a $\JPC=2^{++}$
assignment), and is 783 below that for a fit with the $X(1810)$ omitted.
The latter corresponds to a statistical significance of more than 30$\sigma$.
Different $\JPC$ assignments for the phase space contribution are tested in the PWA fit
and $\JPC$ = 0$^{-+}$ is favored.
The assigned values for the $\JPC$, mass, width and number of events for the five
components for the best fit solution are summarized in Table~\ref{optimalres}.
The mass and width of the $X(1810)$ are obtained to be $M=(1795\pm7$) MeV/$c^2$
and $\Gamma=(95\pm10$) MeV/$c^2$, respectively, where the errors are statistical
only. The contributions of each component of the best solution of the PWA fit
are shown in Fig.~\ref{pwares}(a).
The changes of the log likelihood value $\Delta\mathcal{S}$ and of the number
of degrees of freedom $\Delta ndf$ that occur when a state is dropped from the PWA fit,
as well as the corresponding statistical significance, are also listed in
Table~\ref{optimalres}. The statistical significance of the $f_{2}$(1950),
$f_{0}$(2020) and $\eta$(2225) contributions are $20.4\sigma$, $13.9\sigma$ and 6.4$\sigma$,
respectively.
The reconstruction and final-selection efficiency of the $X(1810)$ is
determined from a weighted phase space MC sample of $\jpsitogwf$, where the weight is
the differential cross section for the measured events calculated
with the magnitudes and phases of the partial amplitudes from the best solution
of the PWA fit. The efficiency is determined to be 6.8\% and the corresponding
branching fraction is
${\cal B}(\jpsi\to\gamma X(1810))\times {\cal B}(X(1810)\to\of)=(2.00\pm0.08)\times10^{-4}$,
where the error is statistical only.


\begin{table*}[htbp]
\centering
\caption{Results from the best PWA fit solution.}
\begin{tabular}{*{8}{c}}\hline\hline
Resonance&~~J$^{PC}$~~&~~M(MeV$/c^2$)~~&~~$\Gamma$(MeV$/c^2$)~~&~~~Events~~~&~~$\Delta$$\mathcal{S}$~~&~~$\Delta ndf$~~&~~Significance\\\hline
$X(1810)$&0$^{++}$&$1795\pm7$&$95\pm10$&$1319\pm52$&783&4&$>30\sigma$\\\hline
f$_{2}$(1950)&2$^{++}$&1944&472&$665\pm40$&211&2&20.4$\sigma$\\\hline
f$_{0}$(2020)&0$^{++}$&1992&442&$715\pm45$&100&2&13.9$\sigma$\\\hline
$\eta(2225)$&0$^{-+}$&2226&185&$70\pm30$&23&2&$6.4\sigma$\\\hline
phase space&0$^{-+}$&--- &--- &$319\pm24$&45&2&9.1$\sigma$\\\hline\hline
\end{tabular}
\label{optimalres}
\end{table*}

The invariant-mass spectra $M$($\kk\ppp$), $M$($\gamma\ppp$), $M$($\gamma\kk$)
and the cos$\theta_{\gamma}$, cos$\theta_{\omega}$,
cos$\theta_{\phi}$, cos$\theta_K$, $\phi_{\phi}$, and $\chi$ angular distributions of the data
and the PWA fit projections with the best solution as well as the different
components are shown in
Fig.~\ref{pwares}.
Here the angles $\theta_{\gamma}$, $\theta_{\omega}$, $\theta_{\phi}$ and $\theta_K$
are the polar angles of the radiative photon in the $\jpsi$ rest frame,
the normal to the $\omega$ decay plane in the $\omega$ system,
$\phi$ meson momentum direction in the $\of$ rest system,
and the kaon from $\phi$ decay in the $\phi$
rest system, respectively; $\phi_{\phi}$ is the azimuthal angle of the
$\phi$ meson in the $\of$ system and $\chi$ is the angle between azimuthal
angles of the normal to the $\omega$ decay plane and the momentum
of a kaon from $\phi$ decay in the $\of$ system.
The PWA fit projection is the sum of the signal events with the best
solution and the background estimated from the weighted events
in the sideband region.

To determine the goodness of fit, a $\chi^2$ is calculated by comparing
the data and fit projection histograms, where $\chi^2$ is defined as~\cite{goodness}
\beq
  \chi^2 = \sum^{N}_{i=1}\frac{(n_i-v_i)^2}{v_i},
\eeq
and $n_i$ and $v_i$ are the number of events for the data and the fit
projections with best solution in the $i^{\rm th}$ bin of each figure,
respectively. The $\chi^2$ and the number of degrees of freedom ($ndf$) for
each mass and angular distribution are shown in Table~\ref{goodness},
where the number of bins is taken as the number of degrees of freedom.
The values of $\chi^2/ndf$ range between 0.62 and 1.70, indicating
reasonable agreement between data and the fit.
\begin{table*}[htbp]
\centering
\caption{Goodness of fit check for the invariant-mass distributions and
 angular distributions shown in Fig.~\ref{pwares}}
\begin{tabular}{*{10}{c}}\hline\hline
~~Variable~~&~~M($\kk\ppp$)~~&~~M($\gamma\ppp$)~~&~~M($\gamma\kk$)~~&
~~~$\theta_{\gamma}$~~~&~~~$\theta_{\omega}$~~~&~~~$\theta_{\phi}$~~~&
~~~$\theta_K$~~~&~~~$\phi_\phi$~~~&~~~$\chi$~~~\\\hline
$\chi^2$&44.4&36.4&42.4&24.2&12.4&28.2&18.2&26.4&51.0\\\hline
$ndf$&40&35&40&20&20&20&20&20&30\\\hline
$\chi^2$/$ndf$&1.11&1.04&1.06&1.21&0.62&1.41&0.91&1.32&1.70\\\hline\hline
\end{tabular}
\label{goodness}
\end{table*}


Additional fits with different assumptions have been carried out to check
the influence on the parameters of the $X(1810)$.
The masses and widths for the $f_{2}$(1950), $f_{0}$(2020) and $\eta(2225)$
are difficult to determine accurately from
this analysis and the achieved accuracy can not compete with the PDG accuracy because of the dominant
$X(1810)$ component; instead they are fixed to their PDG values in the fits.
If we change the masses and widths of these three mesons by one standard
deviation in the fitting, the log likelihood value changes by
$\Delta \mathcal{S} <$ 3 after refitting mass and width
of the $X(1810)$; these values and the branching fraction remain
consistent within the statistical errors. The maximum difference is taken as
a systematic error. An alternative method to test the influence of
the parameters of the three known mesons is to replace the mesons by
states listed in the PDG tables with the same $J^{PC}$ and similar mass.
When the parameters of $f_2$(1950) are replaced with those of the $f_2$(1910),
the log likelihood value increases by $\Delta$$\mathcal{S}$ = 6 after
refitting the mass and width of the $X(1810)$. When the parameters of $f_0$(2020)
are replaced with those of the $f_0$(2100), the log likelihood value increases by $\Delta$$\mathcal{S}$ = 13.
If the $f_0$(2020) is replaced with 0$^{++}$ phase space, the log likelihood
value increases by $\Delta$$\mathcal{S}$ = 9. If the $f_2$(1950) is
replaced with $f_2$(1910) and the $f_0$(2020) is replaced with a 0$^{++}$
phase space, the log likelihood value increases by $\Delta$$\mathcal{S}$ = 10
after refitting. By comparing the log likelihood values, the combination of
$X(1810)$, $f_{2}$(1950), $f_{0}$(2020), $\eta$(2225) and $\JPC$ = 0$^{-+}$
for the phase-space contribution is found to be the best solution, and the
mass, width and branching fraction of $X(1810)$ changes are less than
twice the statistical errors.

The states listed in the PDG tables with mass above the $\wf$ threshold that are
consistent with decaying into $\wf$ under spin-parity constraints, are the $f_2$(1910),
$f_2$(2010), $f_0$(2100), $f_2$(2150), $f_0$(2200), $f_2$(2300) and $f_2$(2340),
etc. Relative to the best solution of the PWA fit, as these resonances
are added in the fit, the log likelihood value improves by 11.7, 8.1,
1.5, 3.2, 1.7, 2.5 and 0.9 after refitting the mass and width of the $X(1810)$, and the statistical
significance of these additional resonances are all less than 5$\sigma$, while
the mass, width, and branching fraction of $X(1810)$ are consistent with those from the
best solution within statistical errors. The maximum difference between the
best fit result and the result with extra states included is taken
as a systematic error. In the best fit solution, phase space is included and approximated as a broad
$J^{PC}=0^{-+}$ resonance. An additional phase-space distribution amplitude with different
$J^{PC}$ was added to test whether the data contain different
$J^{PC}$ phase-space contributions. When the fit is redone
including additional phase-space contributions with $J^{PC}=0^{++}$, $1^{++}$, $2^{++}$,
the log likelihood value improves by 0.1, 3.8 and 3.7, respectively. No
evidence of phase space contributions with different $\JPC$ values is found,
while the $X(1810)$ mass, width, and branching fraction are consistent with
best solution within statistical errors. The maximum differences are taken as
systematic errors.
The BESIII collaboration has observed two new pseudoscalar resonances,
$X(1835)$ in the $\jpsi\to\gamma\etap\pp$ decay process~\cite{x1835} and the
$X(p\bar{p})$ in the $\jpsi\to\gamma p\bar{p}$ decay
process~\cite{xppbar}.
It is interesting to know whether either of these has a $\omega\phi$ decay mode.
Based on the best solution of the PWA fit, new pseudoscalar states with $M=$1836.5MeV/$c^2$,
$\Gamma=$190.1MeV/$c^2$ and $M=$1832MeV/$c^2$, $\Gamma=$76MeV/$c^2$ are added in the fit,
respectively. The log likelihood value improves by 2.2 and 3.5,
and corresponding statistical significance is 1.1$\sigma$ and 1.6$\sigma$, respectively.

Based on the best solution, a more general test is carried out to investigate
the possible contribution from additional resonances not listed by the PDG.
Additional resonances with specified $\JPC$ and width
are included (one at a time) in the fit, with a mass that ranges from low
to high values. The scans are repeated with different widths and
$\JPC$ values. We find that any additional state contribution has a
statistical significance that is less than 5$\sigma$, and the mass, width
and branching fraction of the $X(1810)$ found in this way are consistent
with the best solution.
This method is used to test whether a new resonance/state can be
included in the data and no evidence for a new extra resonance is observed.
The differences in the $X(1810)$ parameters
due to the possible presence of an additional resonance
are not considered in the systematic error determination.

\section{Systematic uncertainty study}\label{Syserr}

For studies of the systematic uncertainties on the PWA-determined mass, width and branching
fraction values for the $X(1810)$, in addition to those discussed above,
the effect of different background determination has also been studied.
To estimate the systematic uncertainty associated with the background determination, the sideband
regions (requirements III and IV) are shifted away from the signal region by 40MeV/$c^2$ and 15MeV/$c^2$
in the $\ppp$ and $\kk$ invariant masses, respectively, the side-band
normalization factors are re-evaluated, and the
PWA fit is redone using the same procedure. The differences from the best
solution are taken as systematic errors.

For the systematic errors on the branching fraction measurement,
there are additional uncertainties from tracking efficiency, particle
identification, photon detection, kinematic fit, as well as the branching
fraction of the intermediate states and the total number of $\jpsi$ events.
\begin{table*}[htbp]
\centering
\caption{Summary of systematic errors}
\begin{tabular}{*{4}{c}}\hline\hline
~~~Source~~~&~~~Mass(MeV$/c^2$)~~~&~~~Width(MeV$/c^2$)~~~&~~~B.R.(\%)~~~\\\hline
Tracking efficiency    &--- &---&~~8.0\\\hline
Particle Identification&--- &---&~~2.0\\\hline
Photon detection       &--- &---&~~3.0\\\hline
Kinematic fit          &--- &---&~~7.0\\\hline
Intermediate branching ratio&---&---&~~1.3\\\hline
$J/\psi$ total number  &--- &---&~~1.2\\\hline
\multirow{2}*{Components in the best fit}         &~+2.0 &+12.8&+19.1\\
                                               &~-5.1 & ~-6.7&-26.1\\\hline
\multirow{2}*{Resonance parameterization}&+12.2&+17.3& ~+1.6\\
                                               &~-1.0 &-33.0&-40.8\\\hline
Background estimation&~+3.0&~-1.0&~-1.4\\\hline
\multirow{2}*{Total}&+12.8&+21.4&+22.3\\
                    &~-5.2&-33.7&-49.8\\\hline\hline
\end{tabular}
\label{syserr}
\end{table*}

The systematic uncertainty associated with the tracking efficiency has
been studied with $\jpsitoppppb$ and $\jpsi\to K^0_{S} K \pi$,
$K^0_{S}\to\pp$ control samples~\cite{trkpidpho}. The difference between data and MC is 2\% per charged pion
and kaon track. Here, 8\% is taken as the systematic error for the
detection efficiency of charged tracks.

The uncertainty due to the kaon particle identification is determined
from studies of a $\jpsitokstk$ control sample~\cite{trkpidpho}. The difference in the
particle identification efficiency between data and MC is 1\% per kaon.
Here, 2\% is taken as systematic error for the identification of two kaons.

The uncertainty due to photon detection efficiency is 1\% per photon, which
is determined from a $\jpsitorhop$ control sample~\cite{trkpidpho}. Here, 3\% is taken as
systematic error for the efficiency of the three photon detection.

To estimate the uncertainty associated with kinematic fit, selected samples of $\psiptoppjpsi$,
$\jpsitokkp$ and $\psiptoppjpsi$, $\jpsitokkpzpz$ events are used to study
efficiency differences between data and MC. Compared to the final states
of the studied channel, the two control samples have exactly the same charged
tracks but one more or one less photon. The efficiency differences between
data and MC are 4.2\% and 7.0\% for the two samples, respectively. Conservatively,
7.0\% is taken as the systematic error associated with the kinematic fit.

For the branching fractions of $\phitokk$, $\omegatoppp$, and $\pztogg$
decays, the uncertainty on these branching fractions listed in the PDG tables~\cite{PDG}
are taken as a systematic uncertainty for our measurement.
The total number of $\jpsi$ events is  $(225.3\pm2.8) \times 10^{6}$,
determined from inclusive $\jpsi$ hadron decays ~\cite{jpsinumber}, with
an uncertainty of 1.2\%.

A summary of all the uncertainties is shown in Table~\ref{syserr}.
The total systematic uncertainty is obtained by summing up all uncertainty
contributions in quadrature. The systematic uncertainties
on the mass and width of the $X(1810)$ are $^{+13}_{-5}$ MeV/$c^2$
and $^{+21}_{-34}$ MeV/$c^2$, respectively, and the relative systematic
error on the product branching fraction is $^{+22}_{-50}$\%.

In the best PWA fit, the threshold enhancement
$X(1810)$ is parameterized by a Breit-Wigner formula with a constant
width. Since the enhancement structure is near the $\of$ threshold,
other decay modes of $X(1810)$ are expected.
To account for this, the Flatt\'{e} formula~\cite{flatte} is used to parameterize the
structure $X(1810)$. We assume the $X(1810)$ mainly decays to $\wf$
and $\kk$ final states, and g$_{\wf}$ and g$_{KK}$ are the coupling constants
to the two modes, respectively. We test two cases, one with g$_{\wf}$ = 1 , g$_{KK}$ = 0
and the other with g$_{\of}$ = 0.5, g$_{KK}$ = 0.5,
the mass and width of the $X(1810)$ shift by $\pm$19 MeV/c$^2$ and $\pm$75 MeV/c$^2$,
respectively, and while the relative change in the product branching ratio is $\pm$65.1\%.
These are considered as a second systematic error due to uncertainty of the model
dependence.

\section{Summary and Discussion}\label{Summary}
We use (225.3$\pm$2.8)$\times$10$^6$ $\jpsi$ events accumulated with the
BESIII detector to study the doubly OZI suppressed decays of $\jpsi\to\gamma\of$,
$\omega\to\ppp$, $\phi\to\kk$. A strong deviation from three-body phase space
for $\jpsi\to\gamma\omega\phi$ near the $\of$ invariant-mass threshold is observed.
Assuming the enhancement is due to the influence of a resonance, the $X(1810)$,
a partial wave analysis with a tensor covariant amplitude determines that
the spin-parity of the $X(1810)$ is 0$^{++}$, and the statistical significance of
the $X(1810)$ is more than 30$\sigma$. The mass and width of the $X(1810)$
are determined to be $M=1795\pm7$(stat)$^{+13}_{-5}$(syst)$\pm$19(mod) MeV/$c^2$ and
$\Gamma=95\pm10$(stat)$^{+21}_{-34}$(syst)$\pm$75(mod) MeV/$c^2$ and the product branching fraction is measured to be
${\cal B}(\jpsi\to\gamma X(1810))\times{\cal B}(X(1810)\to\of)=(2.00\pm0.08$(stat)$^{+0.45}_{-1.00}$(syst)$\pm$1.30(mod))$\times10^{-4}$,
where the first error indicates the statistical error and the second is the
systematical error. These results are consistent within errors with those
from the BESII experiment~\cite{Ablikim:2006}.

The decay $\jpsitogwf$ is a doubly OZI suppressed process that is expected to be suppressed
relative to $\jpsi\to\gamma\omega\omega$ or $\jpsi\to\gamma\phi\phi$ by at least one order of magnitude~\cite{oneorder}.
The anomalous enhancement observed at the $\of$
invariant-mass threshold and the large measured branching fractions
($\sim$1/2 of ${\cal B}(\jpsi\to\gamma\phi\phi)$~\cite{PDG}) are surprising and interesting.
The enhancement is not compatible with being due either
to the $X(1835)$ or the $X(p\bar{p})$, due to the different mass and spin-parity.
The interpretation of the enhancement as being due to
effects of $\of$ final state interactions (FSI) is not excluded in this analysis.
Searches for this structure in different decays modes, e.g. $K^\ast K^\ast$,
$\omega\omega$, etc., and in other production processes, e.g. $\jpsi\to\phi\of$,
$\jpsi\to\omega\of$ etc., are essential to explore the nature of the enhancement,
and gain more insight in the underlying dynamics. The search for other possible states
decaying to $\of$ would also be of interest.
Contributions from $0^{++}$, $0^{-+}$, $2^{++}$ partial waves are found to be necessary
in the PWA fit and simply assigned to the $f_0(2020)$, $\eta(2225)$ and $f_2(1950)$,
respectively, in this analysis, since the PWA fit is not sensitive to those masses and widths.

\section{Acknowledgments}
The BESIII collaboration thanks the staff of BEPCII and the
computing center for their hard efforts. This work is supported ind
part by the Ministry of Science and Technology of China under Contract
No. 2009CB825200;  National Natural Science
Foundation of China (NSFC) under Contracts Nos. 10625524, 10821063,
10825524, 10835001, 10875113, 10935007, 11125525, 10979038, 11005109, 11079030;
Joint Funds of the National Natural Science Foundation of China under Contracts Nos. 11079008, 11179007;
the Chinese Academy of Sciences (CAS)
Large-Scale Scientific Facility Program; CAS under Contracts
Nos. KJCX2-YW-N29, KJCX2-YW-N45; 100 Talents Program of CAS;
Research Fund for the Doctoral Program of Higher Education of China
under Contract No. 20093402120022;
Istituto Nazionale di Fisica Nucleare, Italy; Ministry of Development of Turkey under Contract No. DPT2006K-120470;
U. S. Department of Energy under Contracts Nos. DE-FG02-04ER41291,
DE-FG02-91ER40682, DE-FG02-94ER40823; U.S. National Science Foundation, University of Groningen (RuG)
and the Helmholtzzentrum fuer Schwerionenforschung GmbH (GSI),
Darmstadt; WCU Program of National Research Foundation of Korea
under Contract No. R32-2008-000-10155-0.


\begin{thebibliography}{99}

\bibitem{Ablikim:2006}
  M.~Ablikim {\it et al.} [BES Collaboration],
  Phys.\ Rev.\ Lett. {\bf 96}, 162002 (2006).

\bibitem{oneorder}
K$\ddot{o}$pke L, Wermes N. $\jpsi$ Decays. CERN-EP/88-93, Physics Reports 174 (1989) 67-227: CERN,
CH-1211 Geneva 23, Switzerland.

\bibitem{Bing-An:2006}
  B. A. Li,
  Phys.\ Rev.\ D {\bf 74}, 054017 (2006).


\bibitem{Kung-Ta:2006}
 K. T. Chao, arXiv:hep-ph/0602190.


\bibitem{Bicudo:2007}
  P. Bicudo {\it et al.},
  Eur.\ Phys.\ J.\ C {\bf 52}, 363$-$374 (2007)

\bibitem{Qiang:2006}
  Q. Zhao {\it et al.},
  Phys.\ Rev.\ D {\bf 74}, 114025 (2006).

\bibitem{D.V.:2006}
D. V. Bugg,
J.\ Phys.\ G {\bf 35}, 075005 (2008)

 \bibitem{belle}
 C. Liu {\it et al.} [Belle Collaboration],
 Phys.\ Rev.\ D {\bf 79}, 0701102(R)(2009).



\bibitem{jpsinumber}
M. Ablikim {\it et al.} [BESIII Collaboration],
Chinese Phsyics C 36 (10), 915 (2012).

\bibitem{bes3}
 M.~Ablikim {\it et al.} [BESIII Collaboration],
 Nucl.\ Instrum.\ Meth.\ A {\bf 614}, 345 (2010).

\bibitem{bepc2}
 J.~Z.~Bai {\it et al.} [BES Collaboration],
 Nucl.\ Instrum.\ Meth.\ A {\bf 344}, 319 (1994);
 Nucl.\ Instrum.\ Meth.\ A {\bf 458}, 627 (2001).


\bibitem{geant4}
S.~Agostinelli {\it et al.} [GEANT4 Collaboration],
Nucl.\ Instrum.\ Meth.\ A {\bf 506}, 250 (2003).

\bibitem{boost}
Z. Y. Deng {\it et al.} HEP $\&$ NP {\bf 30}, 371 (2006).

\bibitem{kkmc1}
 S. Jadach, B. F. L. Ward and Z. Was,
 Comp.\ Phys.\ Commu. {\bf 130}, 260 (2000).

\bibitem{kkmc2}
 S. Jadach, B. F. L. Ward and Z. Was,
 Phys.\ Rev.\ D {\bf 63}, 113009 (2001)

\bibitem{besevtgen1}
 K. T. Chao {\it et al.},
 Modern physics A, {\bf 24} No.1 supp. (2009)

\bibitem{besevtgen2}
 R. G. Ping,
 Chin.\ Phys.\ C {\bf 32}, 599 (2008)

\bibitem{PDG}
J. Beringer et al. (Particle Data Group), Phys. Rev. D {\bf 86}, 010001 (2012).

\bibitem{lundcharm}
 J. C. Chen {\it et al.},
 Phys.\ Rev.\ D {\bf 62}, 034003 (2000).

\bibitem{tensor}
 S. Dulat and B. S. Zou,
 Eur.\ Phys.\ J A {\bf 26}, 125 (2005).

\bibitem{FUMILI}
I.Silin,CERN Program library D510, (1971).

\bibitem{goodness}
 M. Ablikim {\it et al}. [BES Collaboration],
 Phys.\ Rev.\ D {\bf 72}, 092002 (2005).

\bibitem{x1835}
M. Ablikim {\it et al}. [BES Collaboration],
Phys.\ Rev.\ Lett. {\bf 106}, 072002 (2011).

\bibitem{xppbar}
M. Ablikim {\it et al}. [BES Collaboration],
Phys.\ Rev.\ Lett. {\bf 108}, 112003 (2012).

\bibitem{flatte}
 S. M. Flatt\'{e},
 Phys.\ Lett.\ B {\bf 63} , 224 (1976)

\bibitem{trkpidpho}
M. Ablikim {\it et al}. [BES Collaboration],
Phys.\ Rev.\ D {\bf 83}, 112005 (2011).



\end{thebibliography}
\end{document}